\documentclass[preprint]{aastex63}

\usepackage{amsmath}

\received{}
\revised{}
\accepted{}
\submitjournal{}

\shorttitle{Heating of Helium ions}
\shortauthors{Shi et al.}

\begin{document}

\title{Proton and Helium Heating by Cascading Turbulence in a Low-beta Plasma}

\correspondingauthor{Zhaodong Shi}
\email{shizd@pmo.ac.cn}

\author{Zhaodong Shi}
\affiliation{Key Laboratory of Dark Matter and Space Astronomy, Purple Mountain Observatory, Chinese Academy of Sciences, Nanjing 210008, Jiangsu, China}
\affiliation{School of Astronomy and Space Science, University of Science and Technology of China, Hefei 230026, Anhui, China}

\author{P. A. Mu\~noz}
\affiliation{Max-Planck-Institute for Solar System Research, 37077 Göttingen, Germany} 
\affiliation{Center for Astronomy and Astrophysics, Berlin Institute of Technology, 10623 Berlin, Germany}

\author{J. B\"uchner}
\affiliation{Max-Planck-Institute for Solar System Research, 37077 Göttingen, Germany}
\affiliation{Center for Astronomy and Astrophysics, Berlin Institute of Technology, 10623 Berlin, Germany}

\author{Siming Liu}
\affiliation{School of Physical Science and Technology, Southwest Jiaotong University, Chengdu 611756, Sichuan, China}



\begin{abstract}
How ions are energized and heated is a fundamental problem in the study of energy dissipation in magnetized plasmas.
In particular, the heating of heavy ions (including ${}^{4}\mathrm{He}^{2+}$, ${}^{3}\mathrm{He}^{2+}$ and others) has been a constant concern for understanding the microphysics of impulsive solar flares.
In this article, via two-dimensional hybrid-kinetic Particle-in-Cell simulations, we study the heating of Helium ions (${}^{4}\mathrm{He}^{2+}$) by turbulence driven by cascading waves launched at large scales from the left-handed polarized
Helium ion cyclotron wave branch of a multi-ion plasma composed of electrons, protons, and Helium ions.
We find significant parallel (to the background magnetic field) heating for both Helium ions and protons due to the formation of  beams and plateaus in their velocity distribution functions along the background magnetic field. 
The heating of Helium ions in the direction perpendicular to the magnetic field starts with a lower rate than that in the parallel direction, but overtakes the parallel heating after a few hundreds of the proton gyro-periods due to cyclotron resonances with mainly obliquely propagating waves induced by the cascade of injected Helium ion cyclotron waves at large scales.
There is however little evidence for proton heating in the perpendicular direction due to the absence of left-handed polarized cyclotron waves near the proton cyclotron frequency.
Our results are useful for understanding the preferential heating of ${}^{3}\mathrm{He}$ and other heavy ions in the ${}^{3}\mathrm{He}$-rich solar energetic particle events, in which Helium ions play a crucial role as a species of background ions regulating the kinetic plasma behavior.

\end{abstract}



\section{Introduction} \label{sec:intro}

Solar energetic particles (SEPs) are routinely generated during solar activity events including solar flares and coronal mass ejections \citep{reames99}. According to their origins \citep{reames13}, SEP events can be classified into two categories: gradual and impulsive events. The latter are produced in impulsive solar flares. Because one of their most distinctive characteristics is the enhancement of the ${}^{3}\mathrm{He}/{}^{4}\mathrm{He}$ ratio, they are also referred to as ${}^{3}\mathrm{He}$-rich events \citep{mason07}. While recent observations have revealed the solar sources of ${}^{3}{\mathrm{He}}$-rich SEPs \citep{bucik20}, it is still under debate what processes are responsible for the preferential heating and acceleration of ${}^{4}\mathrm{He}$ and ${}^{3}\mathrm{He}$ ions, as well as of other heavy ions.

Wave-particle interactions are considered to be one of the most probable mechanisms for generating ${}^{3}\mathrm{He}$-rich SEPs \citep{fisk78}. As \citet{liu-petrosian06} showed, a stochastic acceleration process can consistently account for the acceleration of both ${}^{3}\mathrm{He}^{2+}$ and ${}^{4}\mathrm{He}^{2+}$ observed in ${}^{3}\mathrm{He}$-rich SEP events. This process is based on the resonant wave-particle scattering by parallel propagating waves in a plasma composed of electrons, protons, and ${}^{4}\mathrm{He}^{2+}$ (referred to as Helium ion(s) hereafter as long as there is no ambiguity. For brevity, the kind of plasma is referred to as an electron-proton-Helium plasma.) A key ingredient of their model is that the presence of minor ions ${}^{4}\mathrm{He}^{2+}$ changes the topology of dispersion relations of wave modes, i.e., there are two branches of ion cyclotron waves: a proton cyclotron (PC) wave and a Helium ion cyclotron (HeC) wave (please refer to Figure 1 of their article). As a result, minor ions have a significant impact on the resonant interactions of ${}^{3}\mathrm{He}^{2+}$ and ${}^{4}\mathrm{He}^{2+}$ with the two ion cyclotron wave branches.
These effects still occur even though the abundance of ${}^{4}\mathrm{He}^{2+}$ ions is much smaller than the major protons under solar coronal conditions. 

Recently, based on different kinds of numerical simulations, several scenarios to understand the heating and acceleration of heavy ions in multi-ion plasmas have been proposed.
Most of them are based on plasma turbulence, which is ubiquitous in the solar corona \citep{cranmer2015, zank2021}.
It transfers energy from large fluid scales to small kinetic scales and is thought to play an important role in ion heating.
Considering the typical environments of impulsive solar flares, both \citet{fu-guo20} and \citet{kumar17} have investigated the heating of heavy ions in multi-ion plasmas due to the decaying turbulence initiated by counter-propagating long-wavelength Alfv\'en waves by using 3D hybrid-kinetic and fully-kinetic particle-in-cell (PIC) simulations, respectively.
They found that the turbulence cascaded anisotropically forward to small scales and preferentially transverse to the background magnetic field.
As a result, the heavy ions were preferentially heated perpendicular to the background magnetic field.
While \citet{fu-guo20} argued that the heavy ions were heated via cyclotron resonance with nearly perpendicular magnetosonic waves, \citet{kumar17} claimed that the heavy ions were mainly heated via cyclotron resonance with obliquely propagating Alfv\'en waves.

${}^{3}\mathrm{He}$-rich SEP events have been found to be probably associated with the coronal holes \citep{bucik2018}, in which proton temperature anisotropy has been revealed by observations \citep{cranmer2009}, while a close correlation of the energetic electron beams with these events has been found \citep{wang2012}. Instabilities driven by temperature anisotropy or electron beams are other possible mechanisms of ion heating, by which the ion cyclotron waves can be excited and resonate with ions resulting in heating.
\citet{matsukiyo19} demonstrated that super-Alfv\'enic waves (the low wavenumber part of PC wave) can be excited by the electromagnetic ion cyclotron (EMIC) instability initially driven by a proton temperature anisotropy by using a 1D PIC simulation in an electron-proton-Helium plasma.
They showed that the heavy ions can be firstly pre-accelerated by resonant interactions with proton EMIC waves, and then these pre-accelerated heavy ions can be further accelerated by resonant interactions with super-Alfv\'enic and HeC waves, and thus the energization was mainly transverse to the background magnetic field.

\citet{li2021} applied 1.5D PIC simulations to investigate the heating of heavy ions (including ${}^{4}\mathrm{He}$ and ${}^{3}\mathrm{He}$ ions) in a multi-ion plasma relevant to ${}^{3}\mathrm{He}$-rich SEP events. They injected a relativistic electron beam to start the simulation, and showed that electron and ion cyclotron waves (including PC and HeC waves) can be generated. They demonstrated that the heavy ions were preferentially heated by resonant interactions with the excited cyclotron waves, while ${}^{3}\mathrm{He}^{2+}$ experienced the strongest heating.
They also showed that ions were more efficiently heated for larger magnetization parameter $\omega_{\mathrm{pe}}/\Omega_{\mathrm{ce}}$ and temperature ratio $T_{\mathrm{e}}/T_{\mathrm{p}}$.

Alfv\'en waves are ubiquitous in the solar corona and carry enough energy to probably accelerate the solar wind and perhaps heat the solar corona  \citep{depontieu07, tomczyk2007}, and they may also contribute to the heating of heavy ions in the solar corona and solar wind plasma \citep{chen2018}.
However, large-amplitude Alfv\'en waves are unstable to parametric instabilities \citep{goldstein1978, derby1978}.
Both \citet{araneda09} and \citet{he2016} studied the heating of heavy ions via the parametric instability of a monochromatic Alfv\'en-cyclotron wave and an incoherent Alfv\'en wave spectrum, respectively, by using 1D hybrid-kinetic simulations in electron-proton-Helium plasmas with parameters relevant to the solar wind.
They found that the heavy ions were heated, and the heating perpendicular rather than parallel to the background magnetic field was favored.
While \citet{araneda09} attributed the perpendicular heating to the non-resonant pitch-angle scattering via the pump wave and the excited transverse magnetic fluctuations,
\citet{he2016} argued that the perpendicular heating was due to cyclotron resonance with the excited high-frequency transverse magnetic fluctuations.

Magnetic reconnection is believed to be the ultimate energy source in impulsive solar flares and is likely to be responsible for the heating of heavy ions in SEP events. Both \citet{knizhnik11} and \citet{drake14} investigated the heating and acceleration of ions during magnetic reconnection with a guide field by using 2D PIC simulations in an electron-proton-Helium plasma.
They showed the dominant heating of ions resulted from the pickup behavior of ions during their entry into reconnection exhausts.
This favors the heating transverse rather than parallel to the local magnetic field.
They found that there was a mass-to-charge threshold in pickup behavior which favored the heating of high mass-to-charge ions: for conditions above the threshold, the ions were picked up and became non-adiabatic, 
 and thus a sharp perpendicular heating occurred.
 
 At the larger scales compared to the kinetic scales adopted by the above two studies, \citet{kramolis2022} investigated the acceleration of heavy ions by using a test particle simulation in a background spontaneously fragmenting flare current sheet (SFCS), which was generated in advance by using a 2.5D MHD simulation. They found that a power-law high-energy tail in the ion energy distribution function for each species of ions was obtained due to the first-order Fermi acceleration, and the heavier ions were accelerated preferentially. They claimed that their simulation results were in agreement with the observed SEP events except for the abundance-enhancement factors which were only qualitatively in agreement with the observations.

It is thus still unclear what mechanism can preferentially heat heavy ions under plasma conditions appropriate for impulsive SEPs.
We address this problem by investigating the resonant interaction proposed by \citet{liu-petrosian06} related to the two ion-cyclotron wave branches caused by the presence of heavy minor ions.
This is carried out by means of hybrid-kinetic PIC simulations of collisionless decaying turbulence.
Different from previous studies that initialized the turbulence with long-wavelength Alfv\'en waves, temperature anisotropic distributions, or electron beams, we initially inject a spectrum of HeC waves in order to specifically favor the resonant heavy ion interaction.

Although we focus on the heating of ${}^{4}\mathrm{He}^{2+}$, our results are important to understand the heating and acceleration of ${}^{3}\mathrm{He}^{2+}$ and other heavy ions in ${}^{3}\mathrm{He}$-rich SEP events.
The article is organized as follows: in Section \ref{sec:setup}, we describe the simulation setup including initialization, then in Section \ref{sec:results}, we discuss the results of our simulation including power spectra, dispersion relations, and mainly the heating of Helium ions and protons, and finally, we make conclusions and discussions in Section \ref{sec:conclusion}.

\section{Simulation Setup} \label{sec:setup}

We perform 2D simulations using the hybrid-PIC code CHIEF (Code Hybrid with Inertial Electron Fluid) \citep{munoz18}, which has recently been fully parallelized in order to efficiently run large-scale simulations \citep{Jain2022}. In the hybrid code CHIEF, electrons are treated as an isothermal neutralizing fluid using an EMHD model, while protons and heavy ions are treated as kinetic particles whose distribution functions are advanced in time by solving the corresponding Vlasov equations via the PIC method. A novel feature of CHIEF code is that the inertia of electrons can be taken into account without approximation.
But since the processes to be analyzed here occur at ion time and length scales, we neglect the electron mass in the simulations to be shown.

The simulation domain is a 2D square box in the \textit{yz}-plane (i.e., $L_{y}=L_{z}=L$) with an in-plane homogeneous background magnetic field $\mathbf{B}_{0}=B_{0}\mathbf{e}_{z}$ along the $z$-direction.
This setup allows to have waves with wavenumbers parallel ($k_{\parallel}=k_z$) and perpendicular ($k_{\perp}=k_y$) to the background magnetic field in a 2D geometry.

Both protons and Helium ions are considered.
The abundance of Helium ions is $Y=n_{He,0}/n_{e0}$ = 0.08, and $n_{p0} = n_{e0}-Zn_{He,0}$ due to quasi-neutrality, where $n_{e0}$, $n_{p0}$, and $n_{He,0}$ are the initial number densities of electrons, protons, and Helium ions, respectively, and $Z$ is the charge number of Helium ions.
From now on, the inverse of the proton cyclotron frequency $\Omega_{cp}=eB_{0}/(m_{p}c)$ is chosen as the unit of time, where $c$ is the light speed, $e$ is the proton charge, and $m_{p}$ is the proton mass;
the proton inertial length $d_{p}=v_{A}/\Omega_{cp}$ is chosen as the unit of length, where the proton Alfv\'en velocity is $v_{A}=B_{0}/\sqrt{4\pi m_{p}n_{p0}}$ (in CGS units).
We choose $Z=1$, but the charge-to-mass ratio of Helium ions is $e/(2m_{p})$, i.e., the one half of that of protons.
The simulation domain is square with a side length of $L=192d_{p}$, which is resolved by  $400\times400$ cells.
Due to code constraints, our simulation is actually quasi-2D on the \textit{yz}-plane with 4 cells along the \textit{x}-axis.
We use 960 protons and 200 Helium ions per cell, and periodic boundary conditions are applied for every direction. We have run other simulations varying the number of particles per cell in order to ensure that our chosen values
do not significantly impact the observed results, in particular the heating.
The grid cell size is $\Delta x\approx 0.48 d_{p}$, and the time step $\Delta t=0.025\Omega_{cp}^{-1}$. Electrons, protons and Helium ions have the same initial temperatures, i.e., $T_{e0}=T_{p0}=T_{He,0}$, and the electron plasma beta is $\beta_{e}=0.1$ which is typical of the solar corona plasma, where $\beta_{e}=8\pi n_{e0}k_{B}T_{e0}/B_{0}^{2}$ is the ratio of the electron thermal pressure to the magnetic pressure and $k_{B}$ is Boltzmann constant.

At the beginning of the simulation, the system is perturbed by launching the three longest wavelengths (that the simulation box can accommodate) of left-handed circularly polarized HeC waves in all directions with the form:

\begin{eqnarray}\label{eq:initial_waves}
\delta B_x =\sum_{m_{z}, m_{y}} \varepsilon_{0} B_{0} \cos(m_zk_{0}z+m_yk_{0}y+\varphi_{m_{y},m_{z}}), \\
\delta B_y =\sum_{m_{z}, m_{y}} \varepsilon_{0} B_{0} \sin(m_zk_{0}z+m_yk_{0}y+\varphi_{m_{y},m_{z}}),
\end{eqnarray}

\noindent where the integer $m_{z}$ and $m_{y}$ satisfy $-3 \leq m_{z}, m_{y} \leq 3$ but $m_{z} \neq 0$. This means that a total of 42 wave modes are launched into the system. The minimum wavenumber is $k_{0}=2\pi/L\approx0.033d_{p}^{-1}$, while the maximum is only $k_{\mathrm{max}}=\sqrt{k_{y,max}^{2}+k_{z,max}^{2}} \approx 0.14d_{p}^{-1}$ 
and is still in the fluid-like regime (away from the cyclotron resonance region). $\varphi_{m_{y},m_{z}}$ are random phases.  All waves have the same small initial perturbation amplitude $\varepsilon_{0}\approx 0.0378$, and the rms of initial magnetic field fluctuations $\sqrt{\langle|\delta\mathbf{B}/B_{0}|^{2}\rangle} \approx 0.245$, which is a typical value observed in solar wind turbulence.

Equation \eqref{eq:initial_waves} is general in the sense that it does not indicate a specific wave mode or property other than circular polarization. The specific wave mode is selected from the general expression, which depends on the dispersion relation of such a mode, by considering the initial
bulk velocity perturbations associated with the indicated magnetic field perturbations. HeC wave modes are selected in our case.
For a left-handed circularly polarized wave, the corresponding bulk velocities of protons and Helium ions are given by \citep{araneda09, schreiner14}

\begin{equation} \label{eq:bulkvel}
\frac{\delta\mathbf{u}_{s}}{v_{\mathrm{ph}}}=\frac{-1}{1-\omega_0/\Omega_{\mathrm{c}s}}\frac{\delta\mathbf{B}}{B_{0}} \quad (s=\mathrm{p,\ He}), \quad \text{with } v_{\mathrm{ph}}=\frac{\omega_0}{k_z},
\end{equation}

\noindent where the initial frequency $\omega_{0}$ of every wave is given by the  dispersion relation of left-handed circularly polarized parallel waves (L-mode) in the cold plasma approximation:

\begin{equation} \label{eq:lmode}
\frac{k^{2}}{\omega^{2}}=-\frac{1}{1-ZY}\frac{1}{\omega(\frac{m_{e}}{m_{p}}\omega+1)}-\frac{1}{\omega(\omega-1)}-\frac{ZY}{1-ZY}\frac{1}{\omega(2\omega-1)}.
\end{equation}

\noindent Out of the two wave branches of L-mode for a plasma with Helium ions, we select the HeC wave branch in the expression above, which is the lower frequency branch compared with the PC wave branch. The total initial bulk velocities for protons and Helium ions are the superposition of bulk velocities induced by each wave:

\begin{equation} \label{eq:totbulkvel}
  \delta \mathbf{U}_{s} = \sum_{m_{z}, m_{y}}\delta \mathbf{u}_{s} \quad (s = \mathrm{p,\ He}),
\end{equation}

\noindent where the summation is over all wave modes, and $\delta \mathbf{u}_{s}$ is given by Equation \eqref{eq:bulkvel}.

%
We also write down the right-handed circularly polarized parallel cold plasma waves (R-mode) for reference:

\begin{equation} \label{eq:rmode}
\frac{k^{2}}{\omega^{2}}=-\frac{1}{1-ZY}\frac{1}{\omega(\frac{m_{e}}{m_{p}}\omega-1)}-\frac{1}{\omega(\omega+1)}-\frac{ZY}{1-ZY}\frac{1}{\omega(2\omega+1)}.
\end{equation}

\noindent In Equation \eqref{eq:lmode} and \eqref{eq:rmode}, the wavenumber $k$ and frequency $\omega$ are in units of $d_{p}^{-1}$ and $\Omega_{cp}$, respectively, and the contribution from the displacement current has been ignored. In the MHD limit, the dispersion relation for both L-mode and R-mode represents a standard Alfv\'en wave with a phase speed depending on both ion species:

\begin{equation} \label{eq:alfven}
\frac{k^{2}}{\omega^{2}}=\frac{1+ZY+m_{e}/m_{p}}{1-ZY}=\left(\frac{v_{A}}{v_{A,MHD}}\right)^{2},
\end{equation}

\noindent where $v_{A,MHD}$ is the Alfv\'en velocity in this limit.

Both initial velocity distribution functions of protons and Helium ions are drifting Maxwellians with bulk velocities given by Equation \eqref{eq:totbulkvel}. No perturbations for density, parallel magnetic field, parallel bulk velocities, and electric field are imposed in our simulation.

\section{Simulation Results} \label{sec:results}

Although the considered plasma system initially contains
only HeC waves given by the magnetic and bulk velocity perturbations
Equations \eqref{eq:initial_waves} and \eqref{eq:bulkvel},
other waves will rapidly be excited including but not limited to PC and right-handed waves.
This excitation occurs by turbulent cascading and/or the PIC shot noise.
The latter process is capable to excite all normal plasma wave modes, although with a small amplitude.

During the course of the simulation, the initial long-wavelength waves will cascade forward into a broadband spectrum of kinetic-scale waves. These excited waves can possibly interact with and transfer energy to the heavy ion species. This process is macroscopically quantified as heating.

\subsection{Averaged evolution of the heating}

\begin{figure}
\plotone{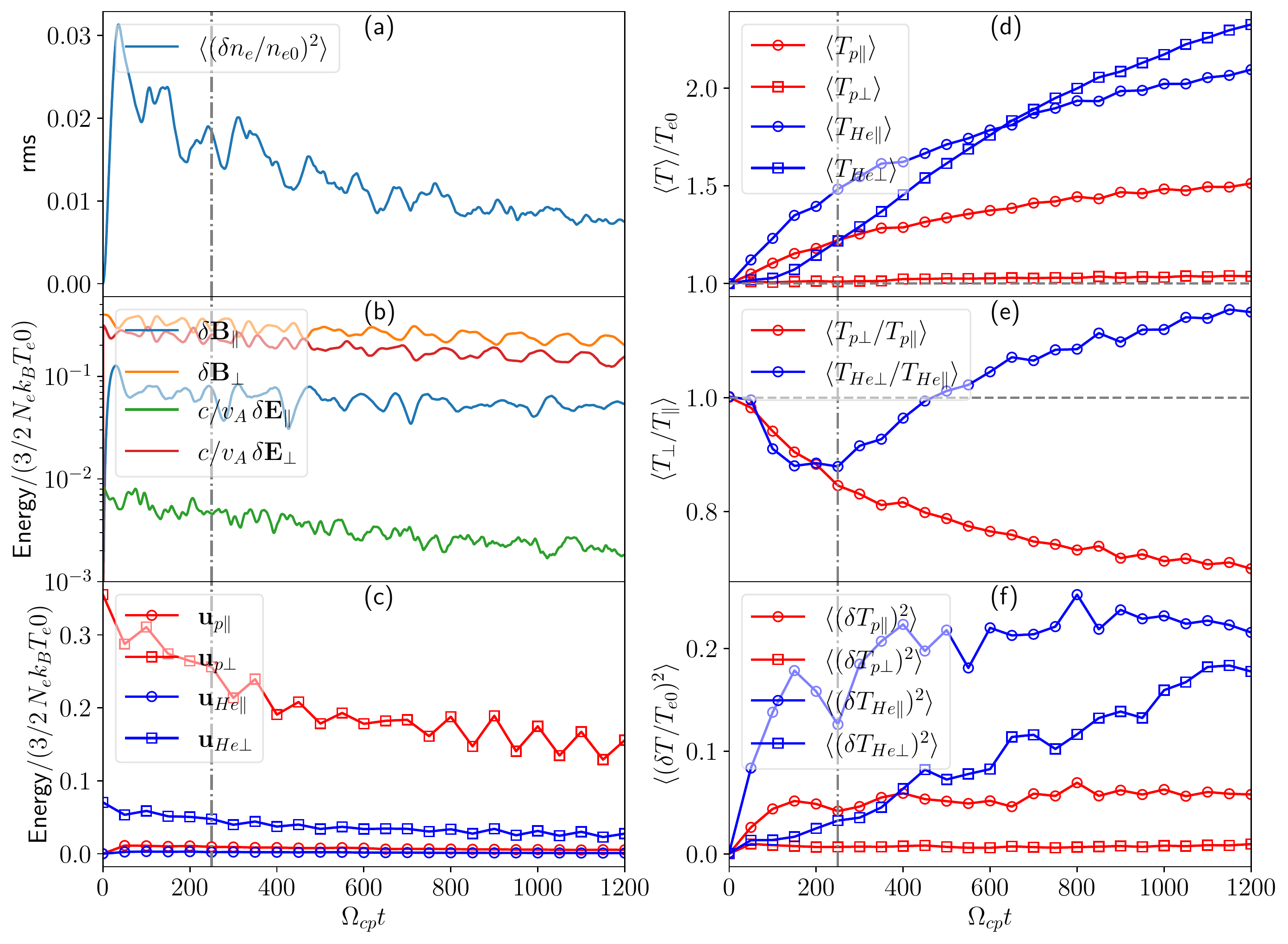}
\caption{The time evolution of (a)  rms of electron number density, (b) energies of parallel and perpendicular magnetic field fluctuations and parallel and perpendicular electric field fluctuations, (c) energies of parallel and perpendicular bulk kinetic energies of protons and Helium ions, (d) box-averaged parallel and perpendicular temperatures of protons and Helium ions, (e) box-averaged temperature anisotropies of protons and Helium ions, and (f) rms values of parallel and perpendicular temperature fluctuations of protons and Helium ions. The vertical dot-dashed line denotes $t\Omega_{cp}=250$. \label{fig:rms}}
\end{figure}

In this subsection, we present the observed heating in our simulation, which we will later connect with resonant wave-particle interactions.

The parallel and perpendicular temperatures relative to the background magnetic field $\mathbf{B}_{0}$ for protons and Helium ions are defined via

\begin{eqnarray}
T_{s\parallel}(\mathbf{x}, t)=\frac{1}{n_{s}}\int |\mathbf{v}_{\parallel}-\mathbf{u}_{s\parallel}|^{2} f_{s}(\mathbf{x}, \mathbf{v}, t)\,d^{3}\mathbf{v}, \\
T_{s\perp}(\mathbf{x}, t)=\frac{1}{2}\frac{1}{n_{s}}\int |\mathbf{v}_{\perp}-\mathbf{u}_{s\perp}|^{2} f_{s}(\mathbf{x}, \mathbf{v}, t)
\,d^{3}\mathbf{v},
\end{eqnarray}

\noindent where $\mathbf{v}_{\parallel}=v_{z}\mathbf{e}_{z}$ and $\mathbf{v}_{\perp}=v_{x}\mathbf{e}_{x}+v_{y}\mathbf{e}_{y}$ are the parallel and perpendicular (microscopic) velocities, respectively.  $f_{s}(\mathbf{x},\mathbf{v},t)$ are the distribution functions of protons ($s$=p) and Helium ions ($s$=He), while their number density and bulk velocity are defined via

\begin{equation}
n_{s}(\mathbf{x},t)=\int f_{s}(\mathbf{x},\mathbf{v},t)\,d^{3}\mathbf{v},\quad
\mathbf{u}_{s}(\mathbf{x},t) = \frac{1}{n_{s}}\int \mathbf{v} f_{s}(\mathbf{x},\mathbf{v},t)\,d^{3}\mathbf{v}.
\end{equation}

Figure \ref{fig:rms}(a) shows the time evolution of the rms value of electron number density fluctuation $\langle(\delta n_e / n_{e0})^{2}\rangle$, where $\langle \cdot \rangle$ means that a quantity ($\cdot$) is averaged over the simulation box.
It decreases gradually and tends to an asymptotic value, which implies that the turbulence is compressible, after the sharp increase at $t\Omega_{cp}\lesssim 50$ because no density perturbation is added at the beginning of the simulation.
Figure \ref{fig:rms}(b) shows the time evolution of the energies of parallel (to the background magnetic field $\mathbf{B}_{0}$) magnetic fluctuation $\int_{V} (8\pi)^{-1} |\delta \mathbf{B}_{\parallel}|^{2}\,dV$, perpendicular magnetic fluctuation $\int_{V} (8\pi)^{-1} |\delta \mathbf{B}_{\perp}|^{2}\,dV$, parallel electric fluctuation $\int_{V} (8\pi)^{-1} |\delta \mathbf{E}_{\parallel}|^{2}\,dV$, and perpendicular electric fluctuation $\int_{V} (8\pi)^{-1} |\delta \mathbf{E}_{\perp}|^{2}\,dV$, where the integral domain $V$ is the whole simulation box. In Figure \ref{fig:rms}(b), the electric fluctuation has been scaled by a factor of $c/v_{A}$.
We can see that the magnetic and electric field energies decrease over the simulation period. The cause of the sharp increases in electric and parallel magnetic fluctuations at $t\Omega_{cp}\lesssim 50$ is the same as that of the sharp increase in density fluctuation.
Meanwhile, the energy of perpendicular fluctuation is much larger than that of parallel fluctuation for both magnetic and electric fields. Figure \ref{fig:rms}(c) shows the time evolution of parallel bulk kinetic energy $\int_{V}
\frac{1}{2} m_{s}n_{s}|\mathbf{u}_{s\parallel}|^{2}dV$ and perpendicular bulk kinetic energy $\int_{V}\frac{1}{2}m_{s}n_{s}|\mathbf{u}_{s\perp}|^{2}dV$ for both protons (red lines) and Helium ions (blue lines).
The perpendicular bulk kinetic energy decreases during the simulation period for both protons and Helium ions, while the parallel bulk kinetic energy is much smaller than the perpendicular one.

The panels (d), (e), and (f) of Figures \ref{fig:rms} show the time evolution of the parallel and perpendicular temperatures, the temperature anisotropy ($T_{s\perp}/T_{s\parallel}$), and the rms values of the parallel and perpendicular temperature fluctuations, respectively, for both protons (red lines) and Helium ions (blue lines).
Figure \ref{fig:rms}(d) shows that the box-averaged parallel and perpendicular temperatures of Helium ions keep increasing over the simulation period.
In contrast, only the box-averaged parallel temperature of protons increases, while the box-averaged perpendicular temperature of protons increases very slightly over the simulation period (at the end of simulation $\langle T_{p\perp}\rangle/T_{p0}\approx1.037$).
The box-averaged parallel temperature is higher than the perpendicular one for Helium ions at $t\Omega_{cp}\lesssim600$, and after that the box-averaged perpendicular temperature surpasses the parallel one.
Figure \ref{fig:rms}(e) shows that the temperature anisotropy (in general $\langle T_{s\perp}/T_{s\parallel} \rangle \neq \langle T_{s\perp} \rangle / \langle T_{s\parallel} \rangle$) of Helium ions keeps increasing after the decrease at the early time, while the temperature anisotropy of protons keeps decreasing.
Figure \ref{fig:rms}(f) shows the time evolution of the rms values of temperature fluctuations ($\delta T_{sl}=T_{sl}-\langle T_{sl}\rangle$ with $l=\,\parallel,\ \perp$). The parallel temperature fluctuation approaches saturation after the early increase for both protons and Helium ions. In contrast, the perpendicular temperature fluctuation of Helium ions keep increasing. The temperature fluctuations of Helium ions attain higher levels than those of protons.
A large temperature fluctuation level implies that the ions are not heated uniformly. In some areas, local temperatures can attain higher values than the corresponding box-averaged temperature, which means that resonant wave-particle interactions are acting strongly in these areas, as we will demonstrate later.


Considering the difference in the heating of the Helium ions at $t\Omega_{cp}\lesssim600$ and at $t\Omega_{cp}\gtrsim600$, in the next subsections, we analyze our results at $t\Omega_{cp}=250$ and at the end $t\Omega_{cp}=1200$ of the simulation.
The time $t\Omega_{cp}=250$ (as shown by the vertical gray dot-dashed line in Figure \ref{fig:rms}) is when the temperature anisotropy of Helium ions reaches its minimum, and after that, the parallel temperature fluctuations of Helium ions and protons approach saturation.

\subsection{Power spectra and turbulence \label{sec:ps}}

\begin{figure}
\plotone{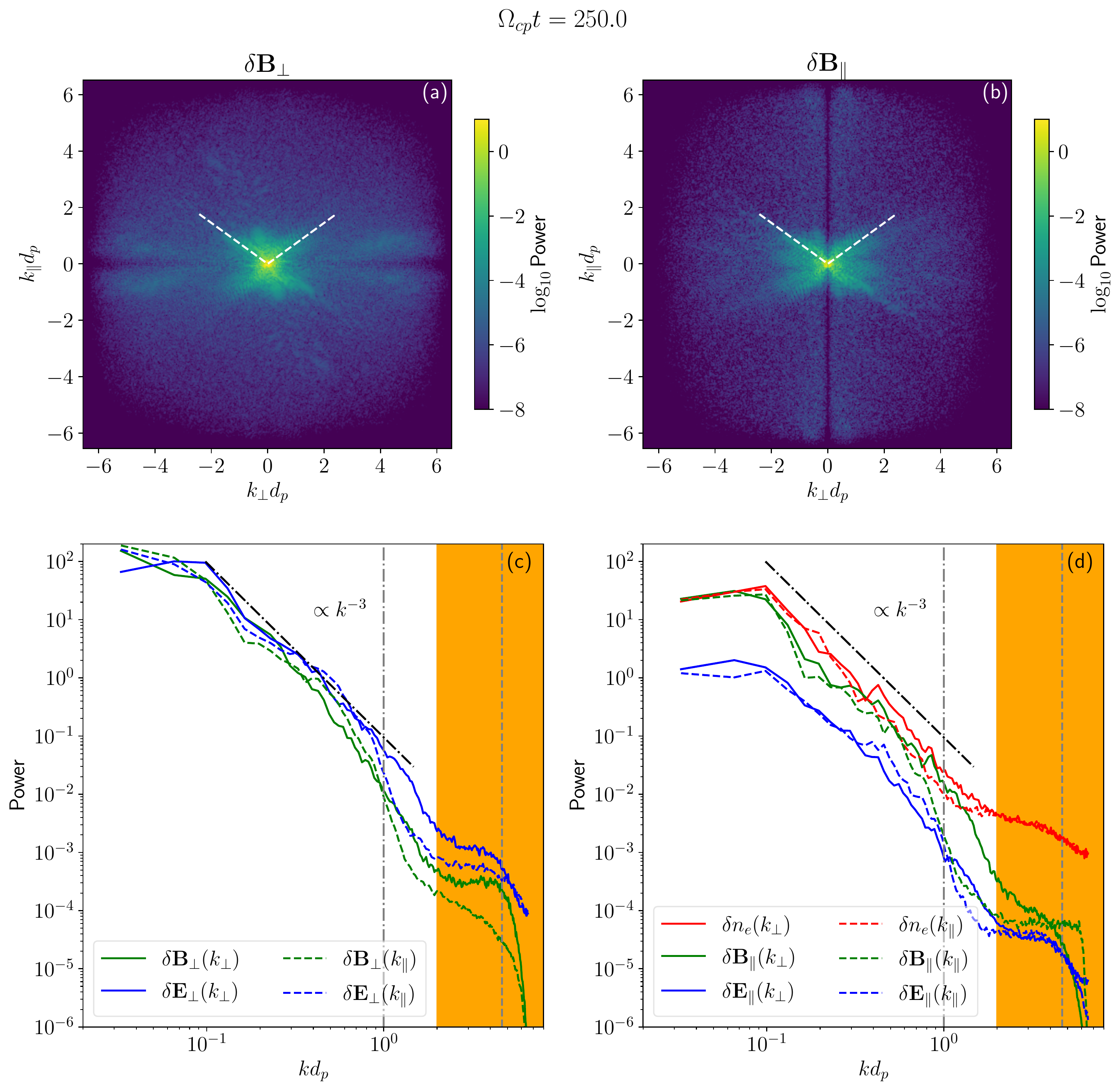}
\caption{Power spectra at $t\Omega_{cp}=250$. (a): 2D power spectrum of $\delta \mathbf{B}_{\perp}$. (b): 2D power spectrum of  $\delta \mathbf{B}_{\parallel}$. (c): 1D power spectra of $\delta \mathbf{B}_{\perp}$ and $\delta \mathbf{E}_{\perp}$. (d): 1D power spectra of $\delta n_{e}$, $\delta \mathbf{B}_{\parallel}$, and $\delta \mathbf{E}_{\parallel}$. The fluctuations $\delta n_{e}$, $\delta\mathbf{B}_{\perp,\parallel}$, and $\delta\mathbf{E}_{\perp,\parallel}$ are normalized in units of $n_{e0}$, $B_{0}$, and $B_{0}v_{A}/c$, respectively. In (c) and (d), the solid lines show the power spectra as the function of perpendicular wavenumber $k_{\perp}$, and the dashed lines show the power spectra as the function of parallel wavenumber $k_{\parallel}$. The dot-dashed black line represents a power-law function with a spectral index of $-3$. The vertical dot-dashed and dashed lines denote that $kd_{p}=1$ and $k\rho_{p}=1$, respectively. The power spectra are affected by the intrinsic noise of the PIC method at small-scales (approximately the orange shaded region).  \label{fig:ps250}}
\end{figure}

\begin{figure}
\plotone{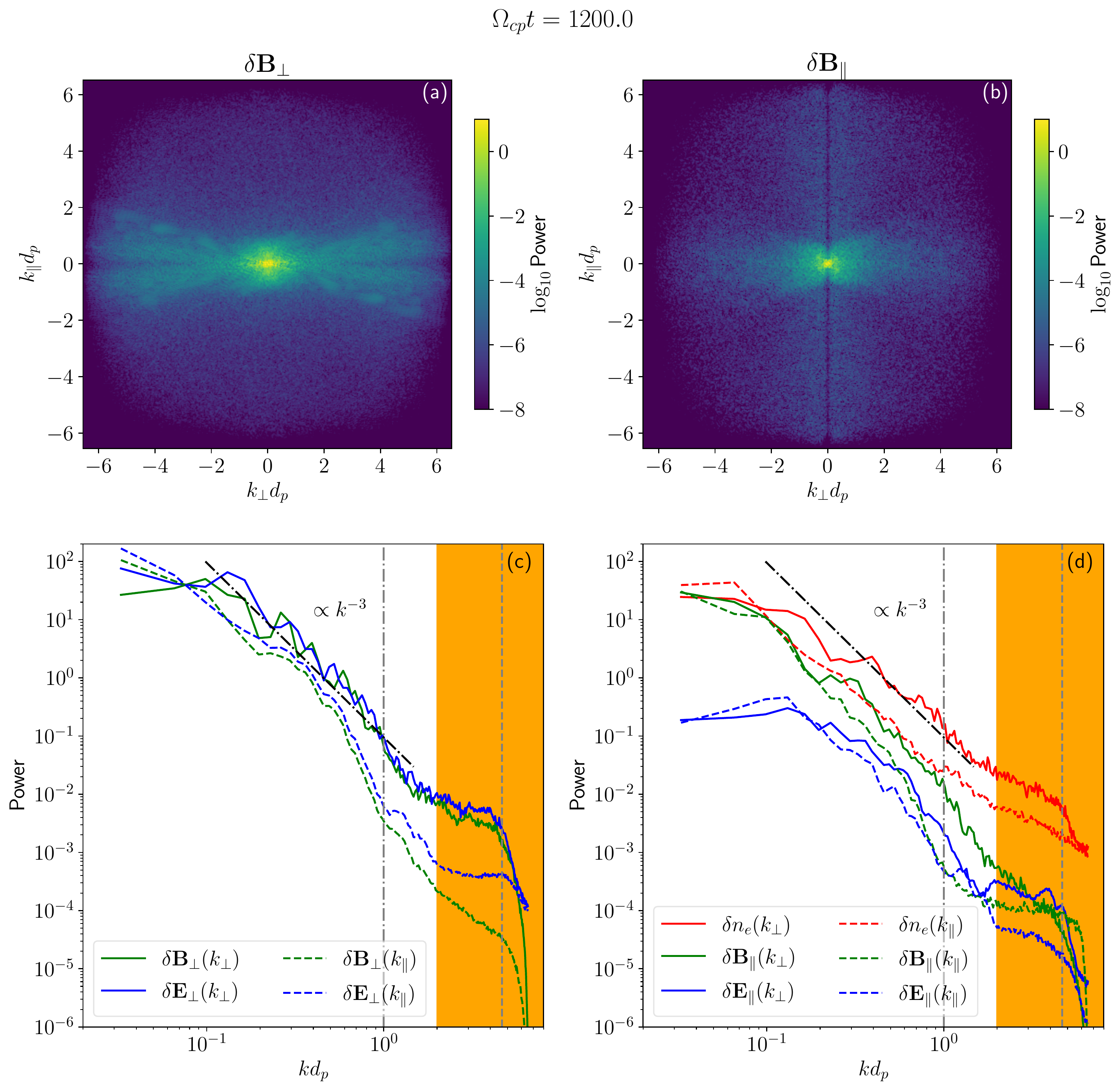}
\caption{Same as Figure \ref{fig:ps250}, but now at $t\Omega_{cp}=1200$. \label{fig:ps1200}}
\end{figure}

In this subsection, we characterize some aspects of the turbulence
developed in the system by means of their power spectra in the wavenumber space.
This allows us to understand some general properties (like the propagation direction)
of the waves that are available
to resonantly interact with the ions and explain the observed heating,
as well as to assess the general properties of the developed turbulence.
We analyze the power spectra of the magnetic field fluctuations ($\delta \mathbf{B}_{\perp}/B_{0}$ and $\delta \mathbf{B}_{\parallel}/B_{0}$), the electric field fluctuations ($\delta \mathbf{E}_{\perp}/(v_{A}B_{0}/c)$ and $\delta \mathbf{E}_{\parallel}/(v_{A}B_{0}/c)$), and the electron number density fluctuation ($\delta n_{e}/n_{e0}$).
All above quantities have been normalized as are given explicitly in the parentheses.

In the top panel of Figure \ref{fig:ps250}, the 2D power spectra in the plane $k_{\parallel}$-$k_{\perp}$ of perpendicular (a) and parallel (b) magnetic fluctuations at $t\Omega_{cp}=250$ are shown.
The magnetic field fluctuations develop mainly along the oblique directions toward the small scales at the early stage, and the dominant propagation direction makes an angle of $\sim 54^{\circ}$ with the background magnetic field, as shown by the white dashed line in panels (a) and (b) of Figure \ref{fig:ps250}.
A similar feature was also observed in the previous simulations of whistler/fast magnetosonic turbulence \citep{svidzinski2009, markovskii2010, markovskii-vasquez2010}.
In those simulations, fast waves were launched as initial conditions\footnote{These authors only considered a plasma composed of electrons and protons, so that the definition of fast and whistler waves is different from our definition (see the next subsection and footnote \ref{ftnt:disp}) as will be adopted for an electron-proton-Helium plasma.}, and magnetic turbulence was observed to develop along the oblique directions.

In Figure \ref{fig:ps250}(c), the 1D power spectra of perpendicular magnetic (green lines) and electric (blue lines) field fluctuations are shown.
The solid lines show the 1D perpendicular power spectra (namely as functions of perpendicular wavenumber $k_{\perp}$), and the dashed lines show the 1D parallel power spectra (namely as functions of parallel wavenumber $k_{\parallel}$).
The power spectra are power-laws with a spectral index of $-3$ for wavenumbers $kd_{p} \lesssim 1$. They steepen at wavenumbers larger than $kd_{p}=1$ (the vertical grey dot-dashed line), and then flatten as wavenumbers approach $k\rho_{p}=1$ (the vertical grey dashed line), where the proton thermal cyclotron radius $\rho_{p}=v_{\mathrm{th}p}/\Omega_{cp}$ while the proton thermal velocity $v_{\mathrm{th}p}=\sqrt{k_{B}T_{p0}/m_{p}}$.
The maximal parallel and perpendicular wavenumber that can be resolved in our simulation is $\pi/\Delta x\approx 6.54d_{p}^{-1}$. At $kd_{p}\gtrsim2$ (the orange shaded region in panels (c) and (d) of Figure \ref{fig:ps250}), the power spectra are affected by the intrinsic noise of the PIC method (the noise level is not shown here, which is estimated using the averaged power spectrum at the very early times for each quantity).
In the fluid regime ($kd_{p}\lesssim1$), there are no significant differences in the spectral energies in the perpendicular and parallel directions for both the perpendicular magnetic and electric field fluctuations.
In the kinetic regime ($1\lesssim kd_{p}\lesssim d_{p}/\rho_{p}$), however, the spectral energy of perpendicular electric fluctuation is larger than that of perpendicular magnetic fluctuation in each direction.
In addition, the kinetic regime also features a spectral energy larger in the perpendicular than in the parallel direction for both the perpendicular magnetic and electric field fluctuations.

Figure \ref{fig:ps250}(d) shows the 1D power spectra of the electron number density fluctuation (red lines), the parallel magnetic (green lines) and electric (blue lines) field fluctuations.
In the fluid regime, the spectral energy of parallel magnetic field fluctuation is comparable to that of electron number density fluctuation, which is also a power-law with a spectral index of $-3$.
For wavenumbers larger than $kd_{p}=1$, however, the power spectrum of the parallel magnetic field fluctuation steepens, while the power spectrum of the electron number density fluctuation flattens.
The spectral energy of the parallel electric field fluctuation is much lower than that of the parallel magnetic field fluctuation in each direction in both fluid and kinetic regimes. The panels (c) and (d) of Figure \ref{fig:ps250} also show that the energies of parallel magnetic and electric field fluctuations are much less than those of perpendicular magnetic and electric field fluctuations, respectively.
This might imply that the heating of ions by Landau resonance is inefficient \citep{li-lu2010}. 

The top panels of Figure \ref{fig:ps1200} show the 2D power spectra of perpendicular (a) and parallel (b) magnetic field fluctuations at $t\Omega_{cp}=1200$.
The power spectra of perpendicular and parallel magnetic field fluctuations are anisotropic in the sense of $k_{\perp} > k_{\parallel}$.
The excited waves are so highly oblique that waves propagating with angles $\vartheta=54^{\circ}\text{ or }126^{\circ}$ are no longer as dominant as at $t\Omega_{cp}=250$.
Figure \ref{fig:ps1200} (c) shows the perpendicular (solid lines) and parallel (dashed lines) power spectra for the perpendicular magnetic (green lines) and electric (blue lines) field fluctuations.
The perpendicular power spectra of perpendicular magnetic and electric field fluctuations in the fluid regime are power-laws with the same index of $-3$ as at $t\Omega_{cp}=250$.
Another similarity between early and later times is that for wavenumbers larger than $kd_{p}=1$, the perpendicular power spectra steepen and then flatten as wavenumbers approach $k\rho_{p}=1$.
Consistent with the scenario of turbulence cascading forward to high perpendicular wavenumbers, the spectral energies in the perpendicular direction of perpendicular magnetic and electric field fluctuations increase significantly in the kinetic regime compared to those at $t\Omega_{cp}=250$.
%
The spectral energy in the parallel direction is much less than that in the perpendicular direction for both perpendicular magnetic and electric field fluctuations.
This is particularly noticeable in the kinetic regime, which implies that turbulence is anisotropic, as expected from turbulence theories and observations \citep{oughton2015, gary2015}.
%
Figure \ref{fig:ps1200}(d) shows that the power spectra of electron number density fluctuation (red lines), parallel magnetic (green lines) and electric (blue lines) field fluctuations.

The anisotropic turbulent cascade has been demonstrated by the previous simulations \citep[e.g.,][]{svidzinski2009, markovskii-vasquez2010, verscharen2012}, consistent with our results.
However, the power-law indices of power spectra are different from each other. For example, \citet{fu-guo20} observed power-law indices of $-2.8$ and $-2$ for the magnetic and electric field fluctuations, respectively, while \citet{kumar17} got a power-law index close to $-5/3$ in the fluid regime.
It is thought that the intensity of the initial magnetic field fluctuation is one of the factors with a significant effect on the spectral indices.
\citet{svidzinski2009} addressed this problem by investigating fast magnetosonic turbulence.
They observed that the more intense the initial magnetic field fluctuation was, the flatter the power spectra of the turbulence were.
We do not discuss further this problem in the present study, since we mainly focus on the heating of ions in the turbulence and are interested in how the anisotropic turbulence influences the heating of ions.

\subsection{Dispersion relations and wave properties \label{sec:dispersion_relations}}

\begin{figure}
\plotone{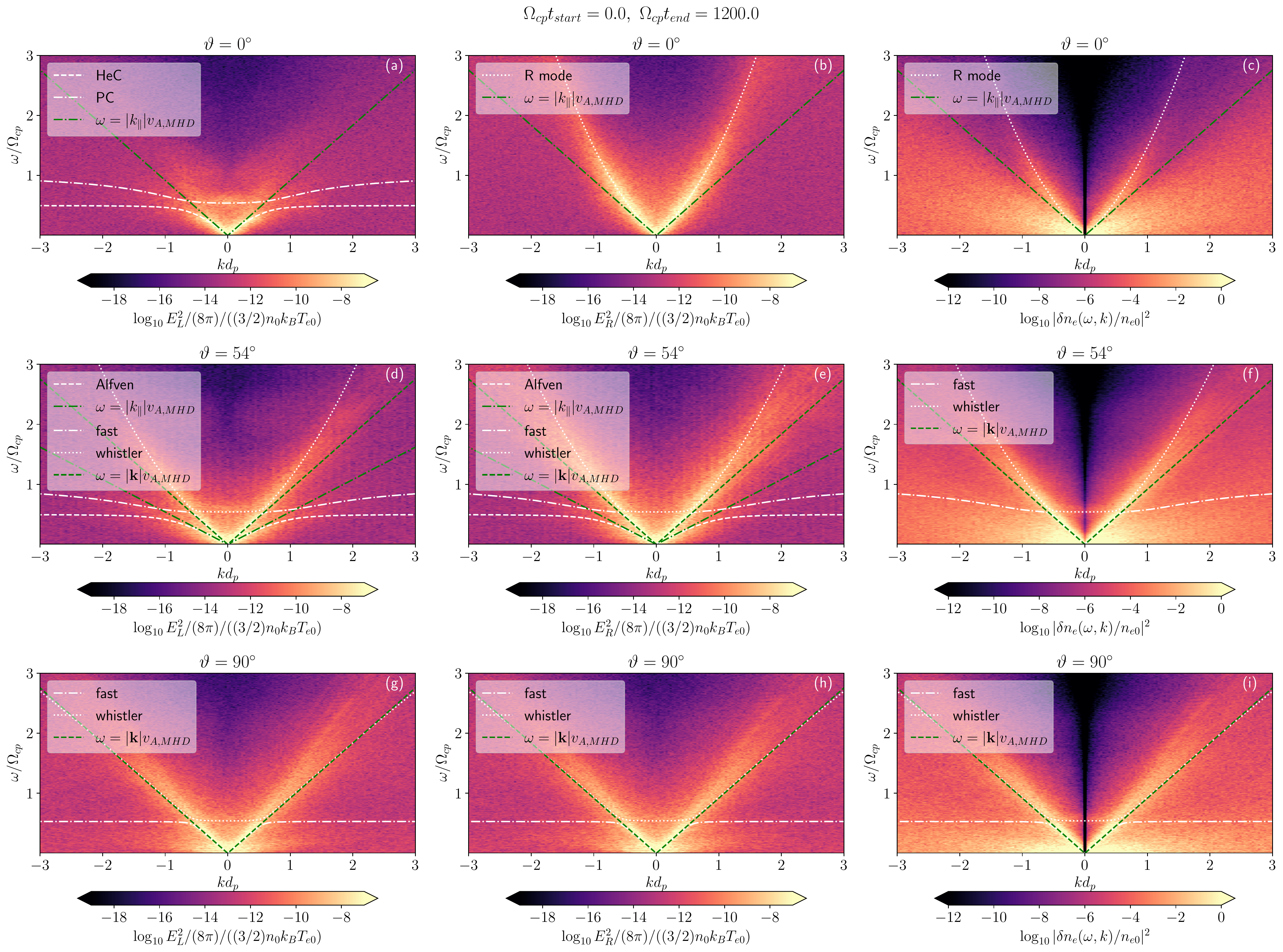}
\caption{The dispersion relations for the L-part $E_{L}$ (a, d, g) and the R-part $E_{R}$ (b, e, h) of electric fields as defined by Eq.~\ref{eq:lrefield}, and the electron number density fluctuation (c, f, i) at $\vartheta = 0^{\circ},\ 54^{\circ},\ 90^{\circ}$. Top panels (a), (b), and (c): White dashed and dot-dashed lines represent HeC and PC branches of L-mode (Equation \eqref{eq:lmode}), respectively. White dotted line represents R-mode (Equation \eqref{eq:rmode}). Middle and bottom panels (d-i): white dashed lines, white dot-dashed lines, and white dotted lines represent the Alfv\'en, fast, and whistler wave branches, respectively. Green dot-dashed lines show the shear Alfv\'en wave ($\omega = |k_{\parallel}|v_{A,MHD}$), and green dashed lines show the fast magnetosonic wave ($\omega = |\mathbf{k}|v_{A,MHD}$). \label{fig:dispersion}}
\end{figure}

In this subsection, we investigate the specific properties of the waves that compose the (wavenumber) power-spectra analyzed in the previous Subsection \ref{sec:ps} by means of their dispersion relations. This allows the identification of the waves developed by the system in the frequency-wavenumber domain, in particular those that can resonate with ions and heat them. Relatively few of the previously mentioned turbulence studies focused on the spectral properties of waves in the frequency-wavenumber domain.

One of the most important wave properties for our purposes is their polarization, since only left-handed polarized waves can resonantly interact with ions effectively. We thus investigate the polarization properties of waves in the dispersion relations of fluctuating electric or magnetic fields via a decomposition of those fields into left- and right-handed parts.
This is performed by means of the following polarization vectors \citep{Zhou2020},

\begin{equation}
\mathbf{e}_{1}=\frac{\mathbf{k}}{|\mathbf{k}|}, \quad
\mathbf{e}_{2}=\mathbf{e}_{x}, \quad
\mathbf{e}_{3}=\mathbf{e}_{1} \times \mathbf{e}_{2},
\end{equation}

\noindent where $\mathbf{e}_{x}$ is the unit vector perpendicular to the \textit{yz}-plane (\{$\mathbf{e}_{x}$, $\mathbf{e}_{y}$, $\mathbf{e}_{z}$\} form a right-handed frame), and $\mathbf{k}=k_{\perp}\mathbf{e}_{y}+k_{\parallel}\mathbf{e}_{z}$ is the wave vector. The circular polarized basis vectors $\mathbf{e}_{L}$ and $\mathbf{e}_{R}$ are defined by

\begin{equation}
\mathbf{e}_{L} = \frac{\mathbf{e}_{2}+\mathrm{i}s\mathbf{e}_{3}}{\sqrt{2}}, \quad \mathbf{e}_{R} = \frac{\mathbf{e}_{2}-\mathrm{i}s\mathbf{e}_{3}}{\sqrt{2}}
\end{equation}

\noindent where the imaginary unit $\mathrm{i} =\sqrt{-1}$, and

\begin{equation}
s = \begin{cases}
\mathrm{sgn} (\omega k_{\perp}), &\quad \text{if }  k_{\parallel}=0, \\
\mathrm{sgn} (\omega k_{\parallel}), &\quad \text{otherwise},
\end{cases}
\end{equation}

\noindent and $\mathrm{sgn}(x)$ is the sign function. Then the left- and right-handed parts of electric fields (being referred to as L-part and R-part of electric fields, respectively) are,

\begin{equation} \label{eq:lrefield}
E_{L}=\mathbf{E}(\omega,\mathbf{k})\cdot\mathbf{e}_{L}, \quad E_{R}=\mathbf{E}(\omega,\mathbf{k})\cdot\mathbf{e}_{R},
\end{equation}

\noindent where $\mathbf{E}(\omega,\mathbf{k})$ is the Fourier transform of electric field $\mathbf{E}(t,\mathbf{x})$ in time domain and real space.

Figure \ref{fig:dispersion} shows the dispersion relations of the L- and R-parts of electric fields (the first and second columns, respectively) and electron number density fluctuation (the third column) for $\vartheta=0^{\circ}$, $54^{\circ}$, and $90^{\circ}$, where $\vartheta$ is the angle between the wave vector $\mathbf{k}$ and the background magnetic field $\mathbf{B}_{0}$. The angle $\vartheta=54^{\circ}$ was identified in panels (a) and (b) of Figure \ref{fig:ps250} as the direction with the strongest spectral power at $t\Omega_{cp}=250$.

The panels (a), (b), and (c) of Figure \ref{fig:dispersion} show the parallel dispersion relations (i.e., at $\vartheta=0^{\circ}$) of the L- and R-parts of electric fields, and electron number density fluctuations, respectively.
Figure \ref{fig:dispersion}(a) shows that HeC and PC branches of L-mode are excited,
although initially only HeC branch is injected.
The HeC branch is confined to almost the interval $|k_{\parallel}|d_{p}\lesssim 0.6$, beyond which it approaches the resonance frequency $\frac{1}{2}\Omega_{cp}$ for this branch and is damped due to the finite temperature effects \citep{ofman05}.
The PC branch of L-mode is also excited,
being confined to almost the same wavenumber interval as the HeC branch.
The PC branch cuts off at $k_{\parallel}=0$, and its cutoff frequency is $\omega_{\mathrm{cut}}\approx(1+ZY)\Omega_{cp}/2=0.54\Omega_{cp}$.
The excited PC branch in our simulation is the super-Alfv\'enic wave defined by \citet{matsukiyo19}.
However, the proton EMIC wave (the higher wavenumber PC wave) is not excited in our simulation, most likely due to the damping as it approaches its resonance frequency $\Omega_{cp}$.
As shown by \citet{matsukiyo19}, this proton EMIC wave can be directly excited by the instability which is driven by proton temperature anisotropy.
In contrast, \citet{li2021} showed that the proton EMIC wave can be excited while the super-Alfv\'enic wave cannot.
These results show that the excitation mechanisms of PC branch by the turbulent cascade and instabilities can be very different and might depend on the specific conditions of each simulation.
Figure \ref{fig:dispersion}(b) shows that R-mode is excited, and Figure \ref{fig:dispersion}(c) shows the dispersion relation of electron number density fluctuation, which follows the dispersion relation of R-mode.

Different from the 1D geometry of \citet{matsukiyo19} and \citet{li2021}, the 2D geometry of our simulation allows us to analyze the dispersion of oblique propagating waves.
The middle panels (d), (e), and (f) of Figure \ref{fig:dispersion} show the dispersion relations of L- and R-parts of electric fields, and electron number density fluctuations, respectively, at $\vartheta=54^{\circ}$ (along the white dashed line in the top panels of Figure \ref{fig:ps250}).
The panels (d) and (e) of Figure \ref{fig:dispersion} show that the Alfv\'en wave branch, fast wave branch, and whistler wave branch\footnote{Here, we use the dispersion relations of waves in a cold electron-proton-Helium plasma to explain our results, and we name the different wave branches following \citet{petrosian2008} (Figure 1 therein) when waves are oblique. There are three branches at the low frequency ($\omega<\sqrt{\Omega_{ce}\Omega_{cp}}$), which are named after Alf\'ven, fast, and whistler waves with increasing frequency at each wavenumber $k$, respectively.
As shown in Figure \ref{fig:dispersion}(d), the dispersion relation of Alfv\'en wave branch is $\omega=|k_{\parallel}|v_{A,MHD}$ (green dot-dashed line) when the wavenumber $|\mathbf{k}|\to 0$, same as the well-known shear Alfv\'en wave in MHD. And the dispersion relation of fast wave branch is $\omega=|\mathbf{k}|v_{A,MHD}$ (green dashed line) when $|\mathbf{k}|\to 0$, same as the fast magnetosonic wave in MHD for a low beta plasma.\label{ftnt:disp}}
are excited,
although the gap between the fast wave branch and the whistler wave branch is too narrow to distinguish them from numerical dispersion relations.
The whistler wave is excited mainly at $|\mathbf{k}|\gtrsim0.6$, while the fast wave is excited mainly at $|\mathbf{k}|\lesssim0.6$.
The spectral power of the excited Alfv\'en wave is much less than that of the other two excited waves. The dispersion relation of electron number density fluctuation from Figure \ref{fig:dispersion} (f) shows that there are two branches: fast and whistler wave branches.
In addition, the whistler wave cuts off approximately
at the frequency $\omega_{\mathrm{cut}}$ when $|\mathbf{k}|=0$, due to the presence of Helium ions.

The bottom panels (g), (h), and (i) of Figure \ref{fig:dispersion} show the perpendicular dispersion relations (i.e., $\vartheta=90^{\circ}$).
Both the fast and whistler wave branches are excited.
However, the Alfv\'en wave branch disappears in this case, since it becomes a zero-frequency mode according to the cold plasma dispersion relation.

From the above, we find that the characteristics of waves in the decaying turbulence can be described very well using the cold plasma wave dispersion in a low-beta electron-proton-Helium plasma.
While the turbulence is compressible, the dispersion of electron number density fluctuations demonstrates that the compressibility is magnetosonic in nature because only the R-mode wave in the parallel propagation and the fast and whistler waves in the oblique propagation are excited.
We will discuss the implications of the excited waves for the heating of ions in the next Subsection \ref{sec:vdfs}, in particular of the obliquely propagating waves.


\subsection{The microphysics of the ion heating \label{sec:vdfs}}

\begin{figure}
\plotone{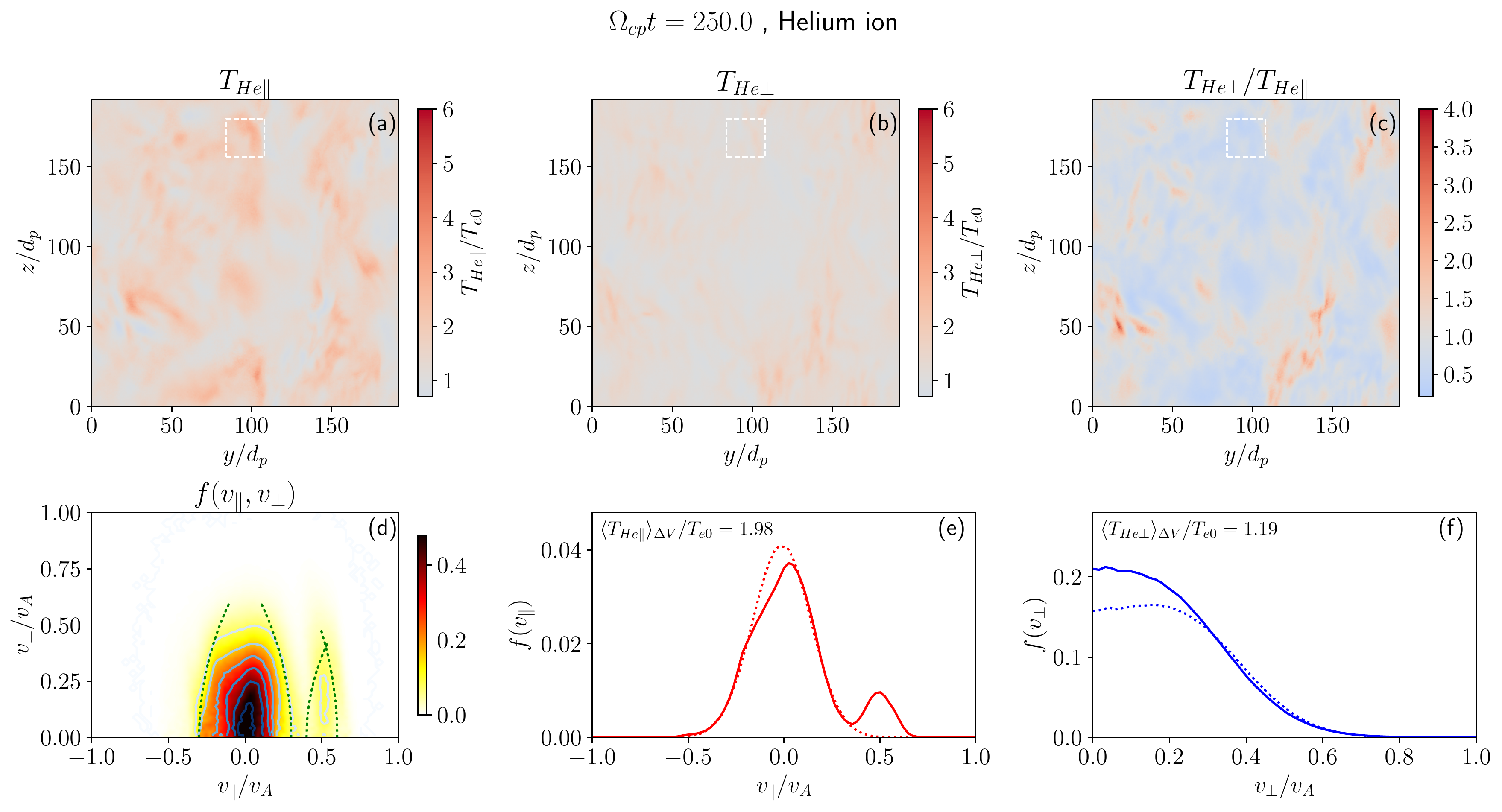}
\caption{(a) the parallel and (b) perpendicular temperatures, and (c) the temperature anisotropy for Helium ions at $t\Omega_{cp}=250$. (d) the 2D VDF $f(v_{\parallel},v_{\perp})$ of Helium ions at $t\Omega_{cp}=250$ in the domain enclosed by a dashed white square in the top panels. (e) and (f) show 1D VDF $f(v_{\parallel})$ and $f(v_{\perp})$, respectively, of Helium ions at $t\Omega_{cp}=250$ (solid lines) and $t\Omega_{cp}=0$ (dashed lines) in the same domain. The upper left annotations in panels (e) and (f) give the parallel and perpendicular temperatures $\langle T_{He\parallel}\rangle_{\Delta V}$ and $\langle T_{He\perp}\rangle_{\Delta V}$, respectively, averaged over the white square at $t\Omega_{cp}=250$. The dotted green curves in (d) are some scattering contours. \label{fig:he4vdf}}
\end{figure}

\begin{figure}
\plotone{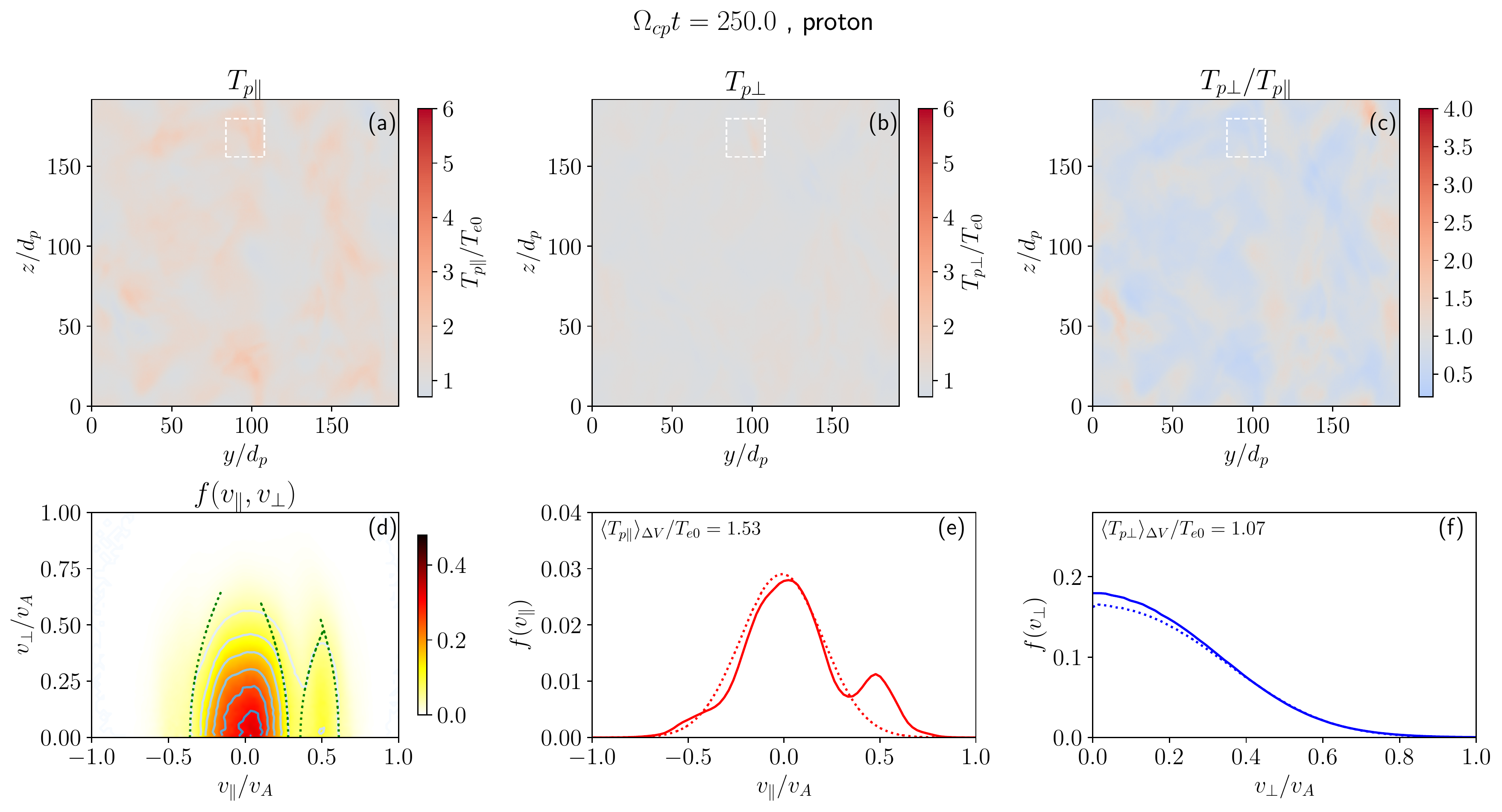}
\caption{(a) the parallel and (b) perpendicular temperatures, and (c) the temperature anisotropy for protons at $t\Omega_{cp}=250$. (d) the 2D VDF $f(v_{\parallel},v_{\perp})$ of protons at $t\Omega_{cp}=250$ in the domain enclosed by a dashed white square in the top panels. (e) and (f) show 1D VDF $f(v_{\parallel})$ and $f(v_{\perp})$, respectively, of protons at $t\Omega_{cp}=250$ (solid lines) and $t\Omega_{cp}=0$ (dashed lines) in the same domain. The upper left annotations in panels (e) and (f) give the parallel and perpendicular temperatures $\langle T_{p\parallel}\rangle_{\Delta V}$ and $\langle T_{p\perp}\rangle_{\Delta V}$, respectively, averaged over the white square at $t\Omega_{cp}=250$. The dotted green curves in (d) are some scattering contours. \label{fig:protonvdf}}
\end{figure}

\begin{figure}
\plotone{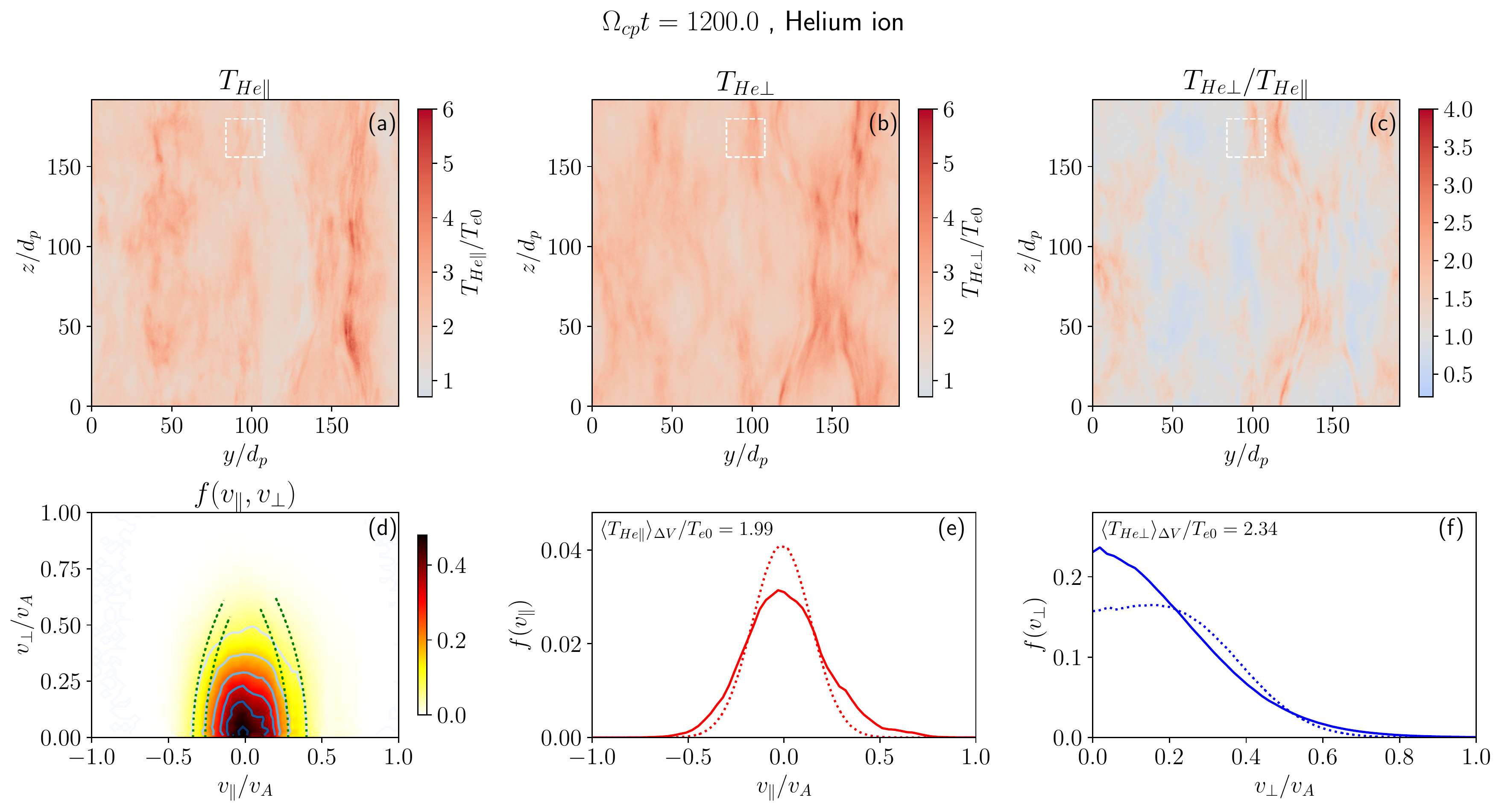}
\caption{Same as Figure \ref{fig:he4vdf}, but now at $t\Omega_{cp}=1200$.  \label{fig:he4vdf1}}
\end{figure}

\begin{figure}
\plotone{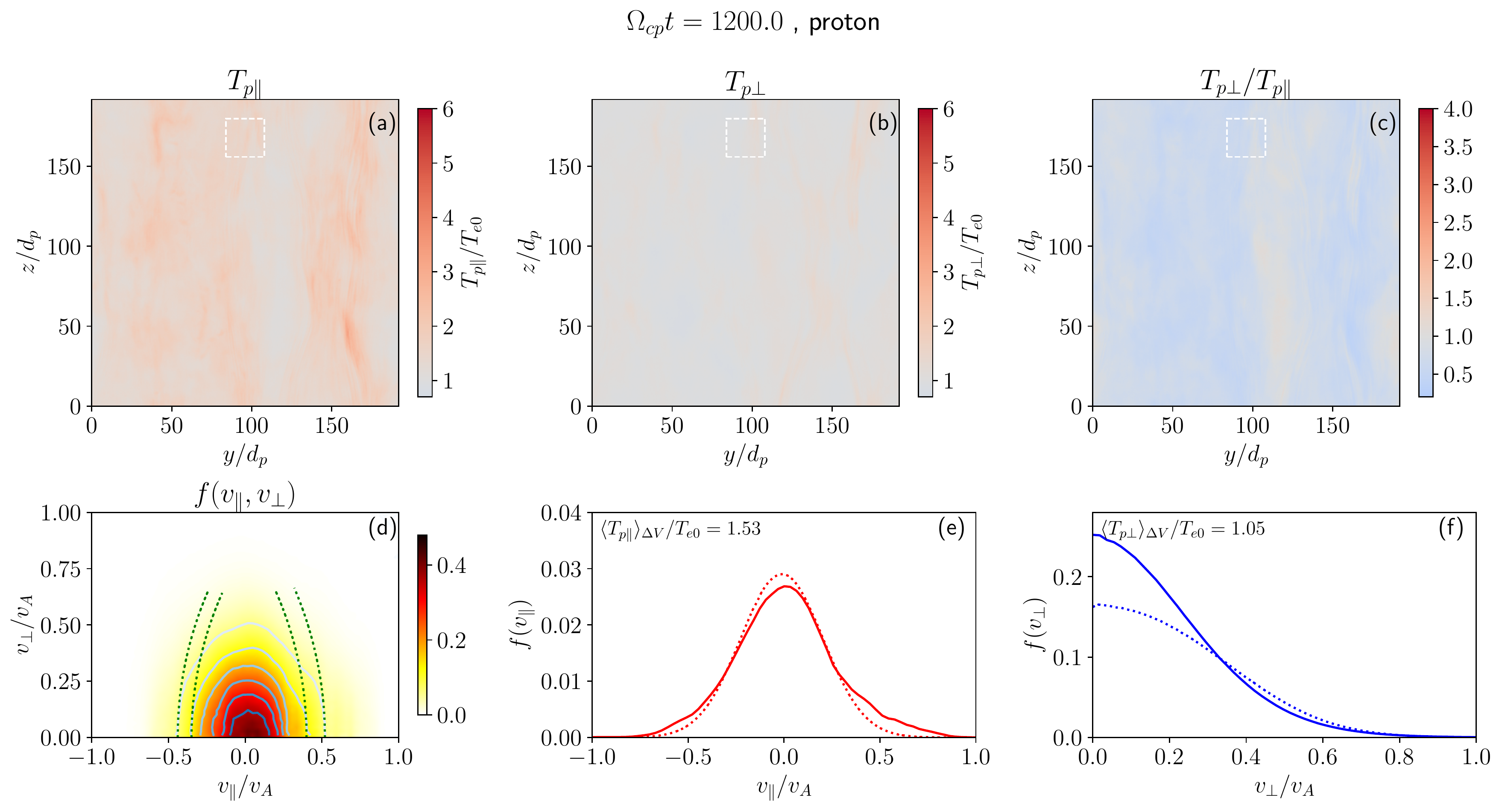}
\caption{Same as Figure \ref{fig:protonvdf}, but now at $t\Omega_{cp}=1200$. \label{fig:protonvdf1}}
\end{figure}



In this subsection, we investigate the heating of Helium ions and protons in detail,
focusing on the interaction between their corresponding distribution functions
and the waves shown in the previous Subsection \ref{sec:dispersion_relations}.

We first define the 2D gyrophase- and volume-averaged velocity distribution function (VDF) in a patch of simulation domain via
\begin{equation}
f_{s}(v_{\parallel},v_{\perp},t)=\frac{1}{\int_{V} f_{s}(\mathbf{x},\mathbf{v},t) d^{3}\mathbf{x}\,d^{3}\mathbf{v}} \int_{\Delta V} d^{3}\mathbf{x}\, \int_{0}^{2\pi} d{\varphi}\, f_{s}(\mathbf{x}, v_{\parallel}, v_{\perp}, \varphi, t) \quad (s=\mathrm{p,\ He}),
\end{equation}

\noindent where the domain $\Delta V$ is the small volume that we are interested in, $f_{s}(\mathbf{x}, v_{\parallel}, v_{\perp}, \varphi, t)=f_{s}(\mathbf{x},\mathbf{v},t)$, $v_{\perp}=\sqrt{v_{x}^{2}+v_{y}^{2}}$, and the gyro phase $\varphi = \arctan (v_{y}/v_{x})$. Likewise, the 1D accumulated volume-averaged VDFs can also be defined: $f_{s}(v_{\parallel},t)=\int f_{s}(v_{\parallel},v_{\perp},t) v_{\perp}\,dv_{\perp}$ and $f_{s}(v_{\perp},t)=\int f_{s}(v_{\parallel},v_{\perp},t) dv_{\parallel}$.

The top panels (a), (b), and (c) of Figure \ref{fig:he4vdf} show the parallel and perpendicular temperatures, and temperature anisotropy, respectively, at $t\Omega_{cp}=250$ for Helium ions.
It is obvious that the Helium ions are heated along the parallel direction in some local areas, while the heating in the perpendicular direction is weaker than that in the parallel direction.
The temperature anisotropy is $T_{He\perp}/T_{He\parallel} < 1$ in most areas of the simulation box, in agreement with Figure \ref{fig:rms}(e), which shows that the box-averaged temperature anisotropy $\langle T_{He\perp}/T_{He\parallel} \rangle < 1$ at $t\Omega_{cp}=250$.

The bottom panels (d), (e), and (f) of Figure \ref{fig:he4vdf} show the 2D VDF $f(v_{\parallel}, v_{\perp})$ and the 1D VDFs $f(v_{\parallel})$ (solid line) and $f(v_{\perp})$ (solid line), respectively, at $t\Omega_{cp}=250$ for Helium ions in the small patch enclosed by the dashed white square (whose area is about $24d_{p}\times 24d_{p}$) as is overplotted in the panels (a), (b), and (c) of Figure \ref{fig:he4vdf}.
The panel (d) of Figure \ref{fig:he4vdf} also shows the contours of the 2D VDF as the solid blue curves. The upper left annotations in panels (e) and (f) of Figure \ref{fig:he4vdf} give the parallel temperature $\langle T_{He\parallel}\rangle_{\Delta V}$ and the perpendicular temperature $\langle T_{He\perp}\rangle_{\Delta V}$, respectively, which are averaged over the white square at $t\Omega_{cp}=250$.
The dashed lines in panels (e) and (f) of Figure \ref{fig:he4vdf} show the initial 1D VDFs $f(v_{\parallel})$ and $f(v_{\perp})$ of Helium ions at $t\Omega_{cp}=0$ for reference, respectively. As shown by panels (d) and (e) of Figure \ref{fig:he4vdf}, an ion beam forms along the parallel direction and is centered on $v_{\parallel} \approx 0.5 v_{A}$.
We attribute the stronger heating in the parallel direction to the formation of the ion beam in the Helium ion VDF in the white square at the early stage.
It is the kinetic heating of the whole ion VDF composed of the ``core" and this beam. The temperarure of the "core" part of the ion VDF does not, however, significantly increase.

The top panels (a), (b), and (c) of Figures \ref{fig:protonvdf} show the parallel and perpendicular temperatures, and temperature anisotropy at $t\Omega_{cp}=250$ for protons, respectively.
The panels (a) and (b) show that protons are heated in the parallel direction in some local areas, while the heating in the perpendicular direction is very weak. This agrees with the global averaged values shown in Figure \ref{fig:rms}(d), which shows only a very small increase in the perpendicular temperature of protons at the end of simulation.
The bottom panels  (d), (e), and (f) of Figure \ref{fig:protonvdf} show the 2D VDF and the 1D VDFs $f(v_{\parallel})$ (solid line) and $f(v_{\perp})$ (solid line), respectively, at $t\Omega_{cp}=250$ for protons in the same small patch enclosed by the white dashed square as in Figure \ref{fig:he4vdf}. The panel (d) of Figures \ref{fig:protonvdf} also shows the contours of 2D VDF as solid blue curves. The dashed lines in panels (e) and (f) of Figures \ref{fig:protonvdf} are the 1D parallel and perpendicular VDFs at $t\Omega_{cp}=0$, respectively.
As shown by panels (d) and (e) of Figure \ref{fig:protonvdf}, the proton VDF features an ion beam which is centered on $v_{\parallel}\approx 0.5v_{A}$. Similar to the Helium-ion case, we also attribute the heating of protons in parallel direction to the formation of the ion beam, although the core proton VDF barely increases its temperature.

Both \citet{araneda09} and \citet{he2016} observed the formation of a proton beam in their simulations, which they attributed to Landau resonance with the ion-acoustic waves (IAWs) excited by a parametric instability.
\citet{perrone11} also observed the formation of a Helium ion beam, in addition to concluding that the formation of ion beams is more efficient for protons than for Helium ions.
In our simulation, although we have observed beam formation for both protons and Helium ions in some locations, the mechanism of beam formation may be different.
As shown in the numerical dispersion in Figure \ref{fig:dispersion}, the density fluctuation is magnetosonic and no IAWs are observed.
Meanwhile, Figure \ref{fig:rms}(b) shows that the energy of parallel electric field $\mathbf{E}_{\parallel}$ is one order of magnitude lower than that of perpendicular electric field $\mathbf{E}_{\perp}$.
Therefore, Landau resonance could be inefficient at forming the beams observed in our simulations.
While \citet{araneda09} show that the proton beam is dynamically stable until the end of their simulation, that is not the case for ours. In our simulation, for the above small volume under consideration, the proton and Helium ion beams forming at the early stage can disappear at the end, as will be discussed later.
This just means that the VDFs are dynamically evolving and do not tend to a stationary state even until the end of the simulation.
The beam formation can be an important reason why the box-averaged parallel temperature is larger than the box-averaged perpendicular temperature for Helium ions at $t\Omega_{cp}\lesssim600$.

The top panels (a), (b), and (c) of Figures \ref{fig:he4vdf1} show that the parallel and perpendicular temperatures, and temperature anisotropy at $t\Omega_{cp}=1200$ for Helium ions, respectively. Helium ions are heated significantly in both the parallel and perpendicular directions, and both the parallel and perpendicular temperatures can reach very high values in some local areas.
As shown in Figure \ref{fig:he4vdf1}, Helium ions are heated stronger in the perpendicular direction than in the parallel direction in most areas of the simulation box.
This agrees with that the box-average temperature anisotropy $\langle T_{He\perp}/T_{He\parallel} \rangle > 1$ according to Figure \ref{fig:rms}(e) at $t\Omega_{cp}=1200$.

The bottom panels (d), (e), and (f) of Figures \ref{fig:he4vdf1} show the 2D VDF $f(v_{\parallel},v_{\perp})$ and the 1D VDFs $f(v_{\parallel})$ and $f(v_{\perp})$, respectively, at $t\Omega_{cp}=1200$ for Helium ions in the same small patch as in the top panels of Figure \ref{fig:he4vdf}.
In this small patch, Helium ions are heated significantly in the perpendicular direction, though the parallel temperature does not increase at $t\Omega_{cp}=1200$ compared to at $t\Omega_{cp}=250$.
Figure \ref{fig:protonvdf1} shows the parallel (a) and perpendicular (b) temperatures, and temperature anisotropy (c), and the 2D VDF (d), and the 1D parallel (e) and perpendicular (f) VDFs at $t\Omega_{cp}=1200$ for protons in the same small patch as in the top panels of Figure \ref{fig:protonvdf}.
Both parallel and perpendicular temperatures of protons do not change at $t\Omega_{cp}=1200$ compared to at $t\Omega_{cp}=250$ in the small patch.
The proton and Helium ion beams disappear at $t\Omega_{cp}=1200$. The Helium ions in the beams might be pitch-angle scattered due to cyclotron resonance, resulting in strong heating in the perpendicular direction and the disappearance of ion beams at the later stage. However, the departures from the Maxwellian for both proton and Helium ion VDFs are still visible.

In our simulation, we find that non-Maxwellian features such as the formation of ion beams and plateaus in the VDFs along the background magnetic field direction are typically observed in those locations where Helium ions and protons are strongly heated in the parallel direction (i.e., the parallel kinetic temperature of the total VDFs increases).
In the next subsection, we demonstrate that non-Maxwellian features in the VDFs of Helium ions and protons imply that resonant wave-particle interactions should play a significant role in ion heating.
%



\subsection{Wave-particle resonances \label{sec:resonance}}

In this subsection we apply the theory of resonant wave-particle interactions to explain the
heating observed in the VDFs of Helium ions and protons, as shown in the previous Subsection \ref{sec:vdfs}.

The wave-particle interactions between waves and ions are strongest when the resonance condition is satisfied, which reads \citep{kennel1966}:

\begin{equation}
\omega -k_{\parallel}v_{\parallel} = n\Omega_{cs}\quad (n=0,\pm1,\pm2,\dots),
\label{eq:resonance}
\end{equation}

\noindent where $k_{\parallel}$ and $v_{\parallel}$ are the wavenumber and ion velocity parallel to the background magnetic field, respectively, and $\Omega_{cs}=q_{s}B_{0}/(m_{s}c)$ is the cyclotron frequency of ion species $s$.

When the waves propagate parallel to the background magnetic field, the harmonic integer is $n=0,\ \pm1$ \citep{tsurutani1997}. 
Landau resonance takes $n=0$ when the waves are electrostatic (longitudinal).
Cyclotron resonances take $n=\pm1$ when the waves are electromagnetic (tranverse).
For positive-charged ions, they are mainly resonant with the L-mode waves via the $n=1$ cyclotron resonance \citep{hollweg-isenberg02}.
Both \citet{matsukiyo19} and \citet{li2021} discussed the $n=1$ cyclotron resonance for protons, Helium ions ($^{4}\mathrm{He}^{2+}$), and ${}^{3}\mathrm{He}^{2+}$ (only in the latter).
They argued that the $n=1$ cyclotron resonance played a dominant role in heating protons and heavy ions, while the heating of heavy ions was preferential \citep{li2021}.
Due to the 1D geometry of their simulations, however, they did not discuss the resonances of ions with obliquely propagating waves.

For obliquely propagating waves, the harmonic number $n$ can take any integer, whereas the $n=1$ cyclotron resonance is the most important for ions.
Both \citet{kumar17} and \citet{fu-guo20} discussed the $n=1$ cyclotron resonance. \citet{kumar17} argued that the heavy ions were heated mainly via the $n=1$ cyclotron resonance with obliquely propagating Alfv\'en waves, while \citet{fu-guo20} argued that the heating of heavy ions is mainly due to resonance with nearly perpendicular magnetosonic waves.
Though they both showed that the turbulence cascaded anisotropically, they did not analyze the numerical dispersion relation, which can be used to identify the available waves for resonances and their propagation directions.

Like \citet{kumar17} and  \citet{fu-guo20}, we also take only the $n=1$ cyclotron resonance into account. The VDF of ion species $s$ tends to a quasi-stationary distribution if the ions diffuse resonantly along the scattering contours defined by \citep{rowlands1966, isenberg-lee96, chandran2010}

\begin{equation}
\frac{1}{2}v_{\parallel}^{2}+\frac{1}{2}v_{\perp}^{2} - \int_{v_{\parallel0}}^{v_{\parallel}}v_{\mathrm{ph}}(v_{\parallel}^{\prime})\,dv_{\parallel}^{\prime} = \mathrm{const.}
\label{eq:diffusionplateau}
\end{equation}

\noindent where the phase speed of the resonant wave $v_{\mathrm{ph}}(v_{\parallel})=\omega/k_{\parallel}$ is given by the resonance condition $\omega-k_{\parallel}v_{\parallel}=\Omega_{cs}$ and the corresponding dispersion relation. For our purpose, it is enough to use the cold plasma dispersion in an electron-proton-Helium plasma. Equation \eqref{eq:diffusionplateau} says that the ions are pitch-angle scattered along the scattering contours 
while their kinetic energies are conserved as measured in the frame comoving at the wave phase speed with the resonant wave.

\begin{figure}
\plotone{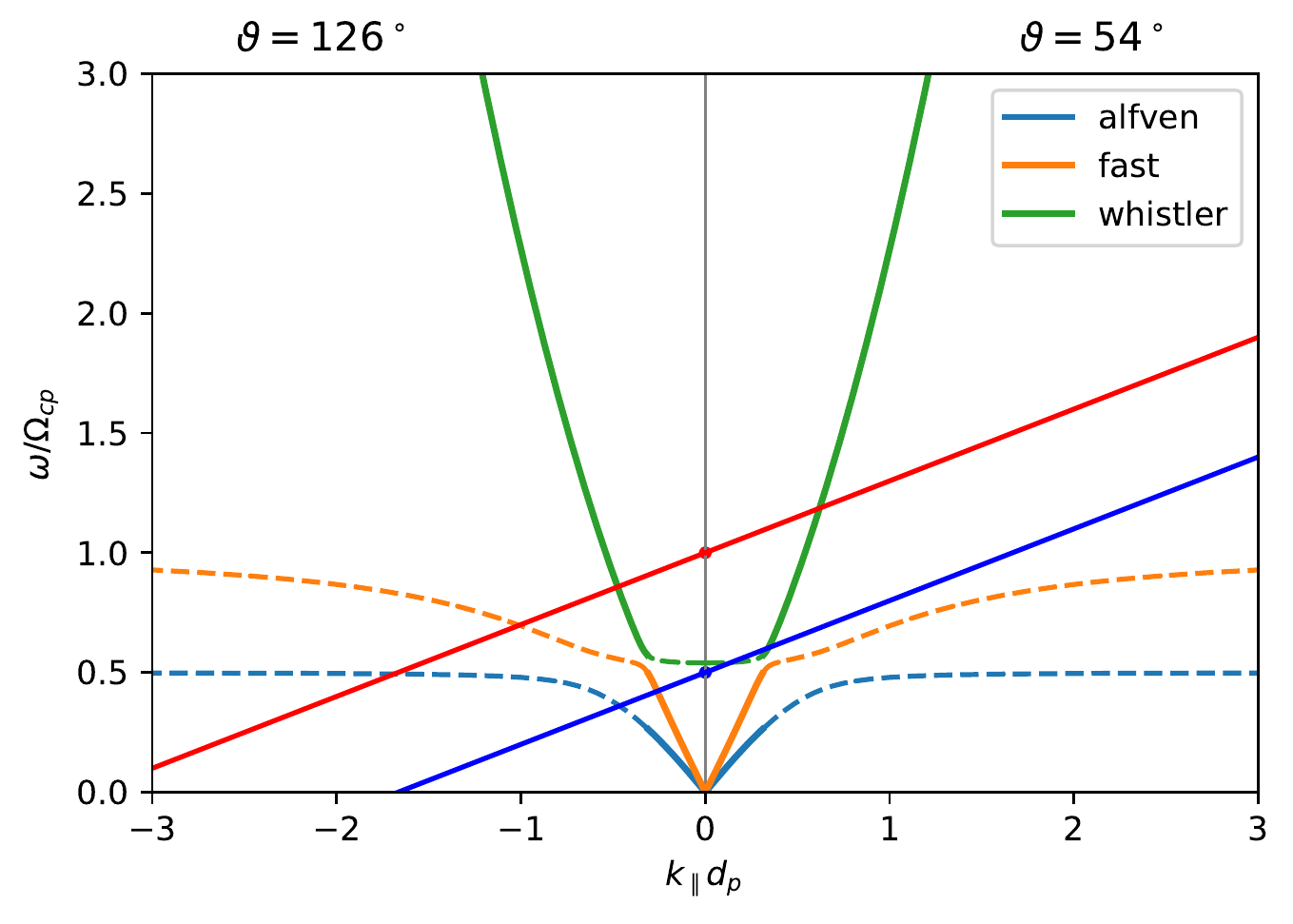}
\caption{The dispersion relation in a cold electron-proton-Helium plasma at propagation angle $\vartheta=54^{\circ}\ (k_{\parallel}>0)$ and $\vartheta=126^{\circ}\ (k_{\parallel}<0)$. The dashed part of each wave branch implicates that the excited wave intensity is weak in the wavenumber range in our simulation. The red and blue lines show the $n=1$ resonance condition $\omega - k_{\parallel}v_{\parallel}=\Omega_{cs}$ for protons and Helium ions, respectively, with the same parallel velocity $v_{\parallel}=0.3v_{A}$.  \label{fig:omkpar}}
\end{figure}

As shown in Subsection \ref{sec:ps}, magnetic field fluctuations develop mainly along oblique directions at the early stage (refer to Figure \ref{fig:ps250}). In addition, due to the finite lower wavenumber range of the excited parallel propagating waves, it is difficult for Helium ions and protons to be resonant with these waves at the early stage in a low-beta plasma.
Therefore, it is reasonable to assume that Helium ions and protons are scattered resonantly mainly by waves with propagation angles $\vartheta=54^{\circ}$ or $\vartheta=126^{\circ}$, as shown in the top panels of Figure \ref{fig:ps250}, at the early stage.
For protons, we consider the case that they are mainly in resonance with whistler waves, while for Helium ions, we consider the case that they are mainly in resonance with whistler and fast waves \citep{li2001, xiong-li2012}.
For making the point clear, Figure \ref{fig:omkpar} shows the dispersion relation of a cold electron-proton-Helium plasma at propagation angles $\vartheta=54^{\circ}\ (k_{\parallel}>0)$ and $\vartheta=126^{\circ}\ (k_{\parallel}<0)$. The dashed part of each wave branch indicates that the corresponding excited wave intensities are weak in our simulation, as shown in the panels (d), (e), and (f) of Figure \ref{fig:dispersion}.
Therefore, we do not consider the resonant interactions of ions with waves in the dashed parts. 
The red and blue lines show the $n=1$ resonance conditions for protons and Helium ions, respectively, with the same parallel velocity $v_{\parallel}=0.3v_{A}$. The protons can resonate simultaneously with the whistler waves at $\vartheta=54^{\circ} \text{ and } 126^{\circ}$.
The helium ions can only resonate with fast waves when their parallel velocities are small, and they can resonate simultaneously with fast and whistler waves when their parallel velocities are larger. 
However, it is hard for protons to be in resonance with Alfv\'en and fast waves, which requires large parallel velocities or wavenumbers, since we are considering a low-beta plasma and the excited wave intensities are weak at large wavenumbers. Similarly, it is also hard for Helium ions to be in resonance simultaneously with Alfv\'en, fast, and whistler waves.

From left to right, the dotted green curves in Figure \ref{fig:he4vdf}(d) represent the scattering contours of Helium ions resonant with the fast ($\vartheta=54^{\circ}$), fast ($\vartheta=126^{\circ}$), whistler ($\vartheta=54^{\circ}$), and fast ($\vartheta=126^{\circ}$) waves, respectively, at $t\Omega_{cp}=250$.
Among these scattering contours, the third is due to the resonance of Helium ions with whistler waves, given that both the resonant wavenumbers and ion parallel velocities are positive.
At the same time, the dotted green curves in Figure \ref{fig:protonvdf}(d) represent the scattering contours of protons resonant with the whistler waves with $\vartheta=54^{\circ},\ 126^{\circ},\ 54^{\circ},\ 126^{\circ}$, respectively.
Though such a simplified model is unlikely to accurately fit the shapes of 2D VDFs of Helium ions and protons, scattering contours can still help us understand many features of the VDFs of Helium ions and protons.
As shown in Figure \ref{fig:he4vdf}(d), the first two scattering contours (counting from left to right) can fit the shape of the Helium ion VDF before it gets flatter at the top. However, the third and fourth scattering contours can only fit the shape of the Helium ion beam distribution more loosely.
Similarly, Figure \ref{fig:protonvdf}(d) shows that the first two scattering contours (counting from left to right) can fit the shape of the proton VDF before it gets flatter at the top.
In addition, the third and fourth scattering contours can remarkably fit the shape of the proton beam distribution, which is unlikely to be a coincidence.
In fact, based on the quasilinear diffusion in the velocity space \citep{kennel1966},
\citet{tu2002} demonstrated that the proton beam distributions observed in the solar wind can form via cyclotron resonant scattering. \citet{tu2002} only considered parallel propagating waves, and a drift speed of Helium ions relative to protons along the background magnetic field was included in their model, which had a significant impact on the resonant wave-particle interactions.
Though it is different from \citet{tu2002} that we mainly consider cyclotron resonances with obliquely propagating waves, the kinetic features of the VDFs of Helium ions and protons imply that the resonant wave-particle interaction should be crucial in heating ions.

At $t\Omega_{cp}=1200$, the dotted green curves in Figure \ref{fig:he4vdf1}(d) show the scattering contours of Helium ions resonant with the fast waves with $\vartheta=54^{\circ}$, $\vartheta=54^{\circ}$, $\vartheta=126^{\circ}$, and $\vartheta=126^{\circ}$, respectively, from left to right. Similarly, the dotted green curves in Figure \ref{fig:protonvdf1}(d) show the scattering contours of protons resonant with the whistler waves with $\vartheta=54^{\circ}$, $\vartheta=54^{\circ}$, $\vartheta=126^{\circ}$, and $\vartheta=126^{\circ}$, respectively. These scattering contours, however, do not well fit the shape of the VDFs of Helium ions or protons. This results from the fact that the waves with $\vartheta=54^{\circ}\text{ or }126^{\circ}$ no longer dominate in spectral power at $t\Omega_{cp}=1200$, as shown in Subsection \ref{sec:ps}.

There are several limitations that should be addressed in the above discussion. Firstly, when discussing resonant wave-particle interactions, we have ignored waves with different propagation angles except $\vartheta=54^{\circ}\text{ and }126^{\circ}$, despite the fact that ions might be resonant with waves of different branches and propagation angles simultaneously.
Secondly, we only consider the $n=1$ cyclotron resonance, which is dominant over the other harmonic resonances when waves are not highly oblique, as is the case at $t\Omega_{cp}=250$. However, the higher harmonic resonances become comparable and important when waves are quasi-perpendicular \citep{steinacker1997, terasawa2012}.
Thirdly, the quasilinear diffusion \citep{kennel1966} predicts that the zeroth order VDF tends to a stationary distribution when its gradient (in velocity space) vanishes along scattering contours.
However, we use scattering contours to understand wave-particle interactions in our simulation at the early stage when the VDFs of Helium ions and protons are dynamically evolving and not stationary.
Despite these limitations, we can conclude that resonant wave-particle interactions play a crucial role in ion heating.

\section{Conclusion} \label{sec:conclusion}

In this article, we performed a 2D hybrid-PIC simulation of collisionless decaying turbulence, which was driven by HeC waves initially injected into the system, in a plasma composed of electrons, protons, and Helium ions. We investigated the heating of Helium ions and protons, and discussed the role of resonant wave-particle interactions in ion heating.
Though Helium ions are minor in abundance in the solar corona and solar wind, they have an important influence on the preferential heating and acceleration of heavy ions in the ${}^{3}\mathrm{He}$-rich SEP events and other kinetic processes in the solar corona and solar wind.

We analyzed the power spectra of the turbulence. We found that the turbulence cascaded along the oblique directions toward the kinetic scales at the early stage. For instance, we analyzed the power spectra in the wavenumber space of the turbulence at $t\Omega_{cp}=250$ and found the dominant directions made angles of $54^{\circ}$ and $126^{\circ}$ with the background magnetic field. Consistent with the previous results \citep{verscharen2012, kumar17, fu-guo20}, later in the simulation, the turbulence cascaded anisotropically in the sense of $k_{\perp}>k_{\parallel}$, and the excited waves were highly oblique at the end.

We also analyzed the dispersion of the excited waves in 3D wavenumber-frequency domain in the turbulence. All the eigenmodes were excited in our simulation, including both parallel and oblique modes. We found that the excited waves can be described very well using the cold plasma approximation in an electron-proton-Helium plasma.
The 2D geometry of our simulation made it possible to analyze the dispersion of waves at oblique propagation angles. In particular, we analyzed the dispersion of waves propagating at angles $\vartheta=54^{\circ}$ $\text{ or }126^{\circ}$, which were found to be the dominant directions in the wavenumber space at the early stage.
This differed from the previous studies in which only parallel waves were taken into account in 1D simulations \citep{matsukiyo19, li2021}.
Though massive parallel 3D kinetic simulations are available now, such a diagnostic seems not to be widely applied to studying plasma turbulence \citep{fu-guo20, kumar17}.

With the above two diagnostics, we analyzed the heating of Helium ions and protons in the turbulence and discussed resonant wave-particle interactions of ions with excited waves.
We found that Helium ions were heated significantly in both the parallel and perpendicular directions.
And the heating of Helium ions has two stages. At the first stage, the box-averaged parallel temperature was higher than the box-averaged perpendicular temperature. At the second stage, however, the box-averaged perpendicular temperature surpassed the box-averaged parallel temperature.
Therefore, perpendicular heating is preferred at the end.
The protons were heated in the parallel direction, while the heating in the perpendicular direction was quite weak.
We found ion beams and/or plateaus in VDFs along the background magnetic field direction for both Helium ions and protons, especially at the early stage. This was manifested as a parallel heating of the corresponding total VDFs.
We analyzed the VDFs of Helium ions and protons in a small volume at $t\Omega_{cp}=250$ in detail.
We found that the resonant wave-particle interactions can explain many features of the VDFs, including the beam formation, by assuming that the ions were resonant with waves at oblique propagation angles $\vartheta=54^{\circ}\text{ or }126^{\circ}$.
Therefore, the resonant wave-particle interactions of ions with obliquely propagating waves played a significant role in ion heating.

The reason why there is almost no perpendicular heating for protons in our simulation may be related to the dimensionality. As \citet{comisel18} have shown, the protons can be heated significantly in the perpendicular direction in a 3D simulation, while there is almost no perpendicular heating in 1D or 2D simulations.
Their results also demonstrated that dimensionality has little influence on the parallel heating of protons.
Recently, \citet{gonzalez20} also addressed the effects of dimensionality on proton heating by investigating the role of a parametric instability.
In addition, we did not study the effects of the intensity of initial magnetic field fluctuations on the turbulent cascade and ion heating. Insted, we focused on investigating ion heating in a specific turbulence instance in the present study.
The investigation of such problems is left to a future publication.

Our results showed that resonant wave-particle interactions are crucial for the heating of ions. The obliquely propagating waves cannot be ignored in considering wave-particle interactions, while theoretical models of heavy ion heating took only the parallel propagating waves into account \citep{liu-petrosian06}.
It is useful for understanding resonant wave-particle interactions by simultaneously diagnosing the power spectra of turbulence and dispersion relations of excited waves.
Our results are useful for understanding the preferential heating of ${}^{3}\mathrm{He}$ and other heavy ions in the ${}^{3}\mathrm{He}$-rich SEP events, in which Helium ions play a crucial role. Future numerical simulations should investigate the heating of ${}^{3}\mathrm{He}$ and other heavy ions and be compared with the observations of impulsive solar flares.

\acknowledgments
Zhaodong Shi is sponsored by China Scholarship Council (CSC No. 201906340076).
We acknowledge the financial support by the German Science Foundation (DFG) via the projects MU-4255/1-1 and BU-777-17-1, by the National Natural Science Foundation of China (NSFC) via grants: U1931204, 11761131007, 12147208, by the National Key Research and Development Program (2018YFA0404203)
 and the International Scholarship Program (G2021166002L) of the Ministry of Science and Technology (MOST) of China. We acknowledge the computational resources provided by the HPC-Cluster of Institut f\"ur Mathematik, Technische Universit\"at Berlin. This study also benefited from discussions within the International
Space Science Institute (ISSI) team on “Origins of 3He-Rich Solar
Energetic Particles.”

%

\vspace{5mm}


\software{
Hybrid-PIC code CHIEF \citep{munoz18}. Python.
         }






\bibliography{draft}{}
\bibliographystyle{aasjournal}



\end{document}